%% file: usenixsecurity2026.tex
\documentclass[letterpaper,twocolumn,10pt]{article}
\usepackage{usenix}
\usepackage{sankey}
\usepackage{algorithm}
\usepackage{algpseudocode}
\usepackage{multirow}
\usepackage{booktabs}
\usepackage{colortbl}
\usepackage[margin=1in]{geometry}
\usepackage{xcolor}
\usepackage{amsmath}
\definecolor{lightgray}{gray}{0.9}
\definecolor{darkgray}{gray}{0.7}
\definecolor{lightgreen}{RGB}{200,255,200}
\definecolor{greenish}{RGB}{180,255,180}
\usepackage[table]{xcolor}
\usepackage{booktabs}
\usepackage{multirow}
\usepackage{graphicx}
\usepackage{rotating}
\usepackage{colortbl}
\usepackage{geometry}
\usepackage{array}
\usepackage{stfloats}
\usepackage{listings}
\usepackage{xcolor} 
\usepackage{booktabs}
\usepackage{enumitem}
\usetikzlibrary{arrows,automata,positioning}
\usepackage{pgfplots}
\usepackage{amsmath}
\usepackage{amssymb}
\usepackage{amsfonts}
\pgfplotsset{compat=1.18}
\usepackage{tcolorbox}
\usepackage{amssymb}
\definecolor{darkgray}{gray}{0.75}
\definecolor{lightgray}{gray}{0.9}
\definecolor{highlight}{RGB}{180,255,180}

\definecolor{catrow}{RGB}{235,243,252}
\definecolor{catrow1}{RGB}{210,225,245}
\definecolor{darkgray}{gray}{0.75}
\definecolor{lightgray}{gray}{0.9}
\definecolor{highlight}{RGB}{180,255,180}
\definecolor{codegreen}{rgb}{0,0.6,0}
\definecolor{codegray}{rgb}{0.5,0.5,0.5}
\definecolor{codepurple}{rgb}{0.58,0,0.82}
\definecolor{backcolour}{rgb}{0.95,0.95,0.92}

\lstdefinestyle{mystyle}{
    backgroundcolor=\color{backcolour},   
    commentstyle=\color{codegreen},
    keywordstyle=\color{magenta},
    numberstyle=\tiny\color{codegray},
    stringstyle=\color{codepurple},
    basicstyle=\ttfamily\footnotesize,
    breakatwhitespace=false,         
    breaklines=true,                 
    captionpos=b,                    
    keepspaces=true,                 
    numbers=left,                    
    numbersep=5pt,                  
    showspaces=false,                
    showstringspaces=false,
    showtabs=false,                  
    tabsize=2
}
\usepackage{tabularx,booktabs,xcolor}
\usepackage{makecell}
\usepackage{multirow}
\usepackage{subcaption}
\usepackage{booktabs,xcolor}
\usepackage{xspace}
\usepackage{url} 
\usepackage{tikz}
\usepackage{amsmath}

\usepackage{filecontents}

\newcommand{\commentout}[1]{}

\begin{document}

\date{}

\title{\Large \bf Low-ASR Backdoors: Exploiting Attack Success Rate Reduction and Attacker–Defender Asymmetry}

\author{
    Arham Riaz \qquad Ting Yu\\
    \textit{Mohamed bin Zayed University of Artificial Intelligence}
}

\maketitle

\input{sections/0_abstract}
\input{sections/1_introduction}
\input{sections/2_background_and_motivation}

\input{sections/3_low_asr_backdoor}
\input{sections/4_exp_and_results}

\input{sections/5_discussion}
\input{sections/6_related_work}
\input{sections/7_conclusion}

\cleardoublepage


\cleardoublepage
\bibliographystyle{plainurl}
\bibliography{sample}

\newpage
\clearpage
\appendix
\input{appendix/0_appendix}

\input{appendix/1_appendix}

\input{appendix/2_appendix}
\end{document}

%% file: sections/0_abstract.tex
\begin{abstract}
Backdoor attacks are among the most effective and stealthy attacks in deep learning. Existing attacks and defenses are largely designed and evaluated under the assumption that successful backdoors exhibit high Attack Success Rates (ASRs). In this paper, we show that this assumption creates a fundamental weakness in existing defense paradigms. ASR is not an intrinsic property of a backdoor; rather, it is an attacker-controlled variable that can be deliberately reduced without eliminating the underlying backdoor behavior. We introduce a reverse-training framework that weakens the trigger--target association, producing low-ASR backdoor models while preserving clean-input performance. Through extensive evaluation across multiple datasets, diverse attack families, and multiple architectures, we show that state-of-the-art defenses fail consistently under low-ASR conditions, exposing a fundamental attacker–defender asymmetry.

\end{abstract}

%% file: sections/1_introduction.tex
\section{Introduction}
\label{sec:introduction}

Deep neural networks are increasingly deployed in security- and safety-critical applications, including autonomous driving~\cite{bojarski2016end}, biometric authentication~\cite{jain2011face}, healthcare~\cite{esteva2017dermatologist,litjens2017survey}, financial decision-making~\cite{mohsin2025explaining}, video understanding~\cite{doan2024video}, and large language models~\cite{yan2024llm,zhang2024instruction}. As these systems become integrated into real-world infrastructure, their security has become an important concern. Prior work has shown that deep learning models are vulnerable to a broad range of attacks, including training-time backdoor attacks~\cite{barreno2010security,li2022backdoor}.

\begin{figure}[t]
    \centering
    \includegraphics[width=\linewidth]{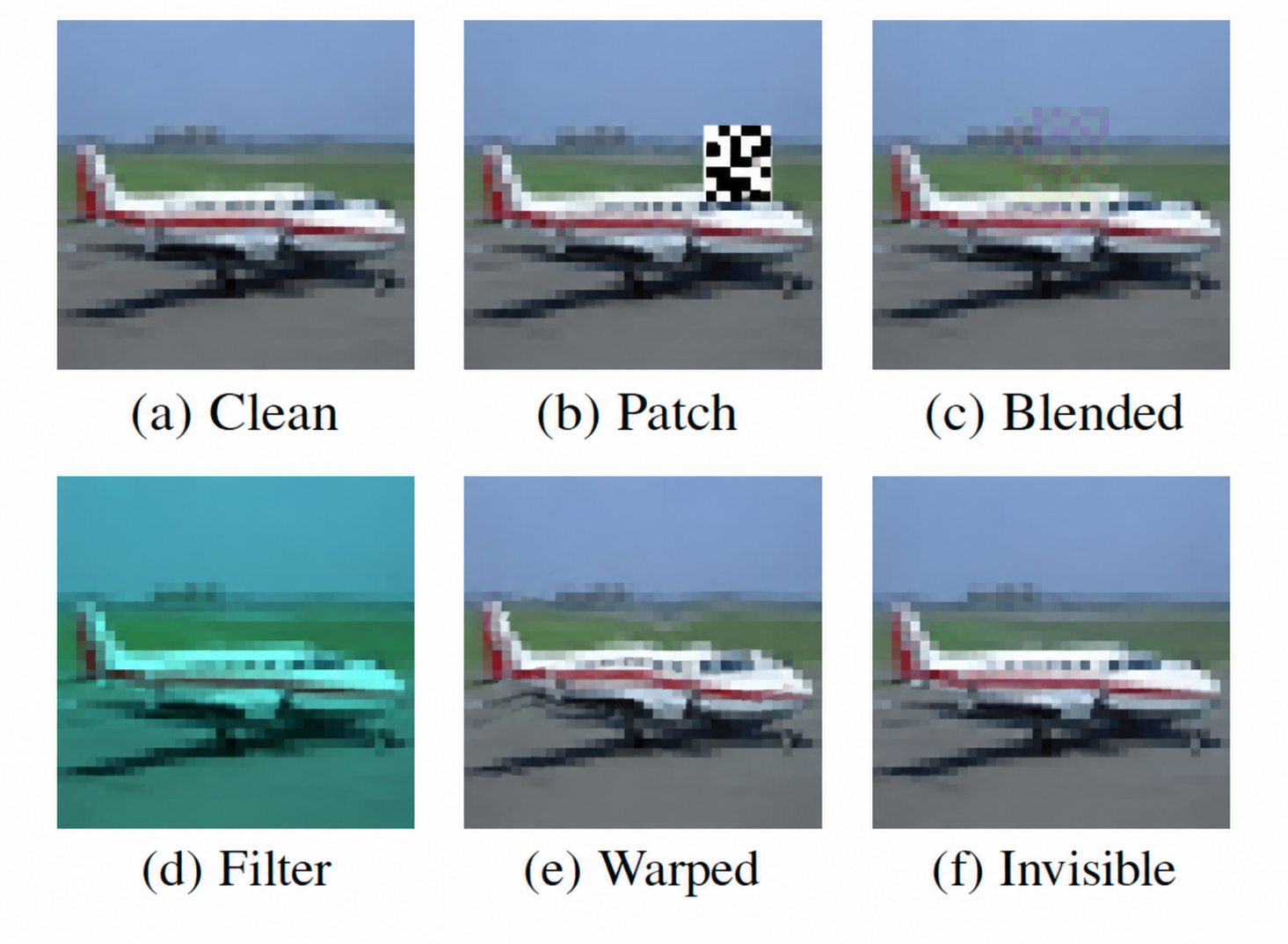}
    \caption{Examples of patch-based, blending-based, filter-based, warping-based, and learning-based backdoor triggers applied to a clean image.}
    \label{fig:trigger_family}
\end{figure}

Backdoor attacks embed attacker-specified behavior into a model while preserving normal performance on clean inputs. Early attacks relied on visible patch triggers~\cite{gu2017badnets}, while subsequent work introduced increasingly stealthy mechanisms, including blended~\cite{chen2017targeted}, reflection-based~\cite{liu2020reflection}, physical~\cite{wenger2021backdoor}, semantic~\cite{bagdasaryan2020backdoor}, compression-resistant~\cite{xue2022compression}, warping-based~\cite{nguyen2021wanet}, and input-dependent triggers~\cite{doan2021lira,salem2022dynamic}. Figure~\ref{fig:trigger_family} illustrates representative examples of these trigger families. In parallel, defenses have been developed using trigger reconstruction~\cite{wang2019neural,tao2022better}, entropy analysis~\cite{gao2019strip}, representation inspection~\cite{fu2023freeeagle,liu2019abs}, poisoned-sample identification~\cite{chen2018detecting,tran2018spectral,qi2023proactive}, and optimization-based trigger recovery~\cite{guo2021aeva,dong2021black,popovic2025debackdoor}.

Despite these advances, attack and defense evaluations commonly assume that an effective backdoor produces a strong and consistent trigger-induced response. Conventional attacks are typically optimized to achieve high Attack Success Rates (ASRs), and defenses are consequently evaluated under conditions in which malicious behavior is highly observable. Widely studied attacks often report ASRs exceeding 90\%~\cite{gu2017badnets,chen2017targeted,nguyen2021wanet,doan2021lira}. This evaluation paradigm leaves an important question underexplored: how reliable are existing defenses when an attacker deliberately reduces ASR?

ASR measures the proportion of triggered inputs classified as the attacker-specified target label. ASR is commonly treated as a measure of backdoor effectiveness. However, ASR is an attacker controlled variable. An attacker can therefore deliberately reduce ASR, trading consistent trigger activation for a weaker and less observable backdoor signal. Consequently, the absence of a strong trigger-induced response does not necessarily imply the absence of an underlying backdoor.

In this work, we study low-ASR backdoors, where the attacker deliberately reduces the backdoor's activation rate while preserving clean-input performance. We investigate whether such backdoors can evade existing defenses while retaining attacker-useful behavior.

This setting creates an \emph{attacker--defender asymmetry}. The first asymmetry concerns the observable signal: the defender requires sufficiently consistent evidence to distinguish a backdoored model from a clean one, whereas the attacker can deliberately weaken that evidence by reducing ASR. The second concerns candidate selection: the defender evaluates a limited set of probes, whereas the attacker can search for an input that induces the desired target prediction. Thus, the defender requires reliable aggregate evidence, while the attacker needs to identify only one successful candidate.

To investigate this setting, we introduce a reverse-training framework for generating controlled low-ASR models. Starting from a conventional high-ASR backdoor model, reverse training applies the original trigger transformation while assigning triggered inputs their ground-truth labels. This process progressively weakens the trigger--target association while maintaining the original trigger, target label, architecture, and attack configuration. The resulting checkpoints allow us to examine how decreasing ASR affects backdoor representations, defense performance, and attacker utility.

We evaluate the framework across MNIST, CIFAR-10, and GTSRB using patch-based, blended, WaNet, and LIRA attacks. We further examine four representative defenses spanning different detection paradigms: Neural Cleanse, STRIP, FreeEagle, and DeBackdoor. Our results show that their detection signals weaken substantially as ASR decreases, causing low-ASR backdoor models to be classified as benign under the evaluated settings. At the same time, gradient-based activation experiments show that triggered inputs evaluated on low-ASR models require substantially fewer optimization steps to reach the attacker-specified target than inputs evaluated on clean models. These findings demonstrate that low observable ASR does not necessarily imply the absence of attacker-useful target-class bias.

In summary, this paper makes the following contributions:

\begin{itemize}

    \item \textbf{Low-ASR Threat Model.}
    We formalize a threat model in which a malicious model supplier deliberately reduces trigger-induced behavior and uses a retained local model to identify successful input candidates before deployment-time submission.

    \item \textbf{Controlled ASR Reduction.}
    We introduce a reverse-training framework that transforms conventional high-ASR backdoors into controlled low-ASR variants while preserving clean-input performance and the original attack configuration.

    \item \textbf{Defense Evaluation.}
    We systematically evaluate Neural Cleanse, STRIP, FreeEagle, and DeBackdoor and show that their detection effectiveness degrades as the observable trigger-induced signal weakens.

    \item \textbf{Backdoor Activation at Low ASR}
    We show that low-ASR models retain measurable target-class bias: triggered inputs require fewer gradient-based optimization steps to reach the attacker-specified target than comparable inputs evaluated on clean models.

\end{itemize}

%% file: sections/2_background_and_motivation.tex
\section{Background}
\label{sec:background}

\subsection{Backdoor Attacks}
\label{sec:background_attacks}

Deep models are vulnerable to backdoor attacks~\cite{barreno2010security,li2022backdoor}. Backdoor attacks are training-time attacks in which an adversary implants hidden malicious behavior into a machine learning model. The attacker associates a trigger $\Delta$ with an attacker-defined target label function $\phi(\cdot)$. Once the backdoor has been implanted in the model, clean inputs are classified normally, while inputs containing the trigger are mapped to attacker-chosen predictions.

Formally, given an input-label pair $(x,y)$, a backdoor model satisfies
\begin{equation}
f_\theta(x) = y,\qquad
f_\theta(x+\Delta) = \phi(y),
\label{eq:backdoor_definition}
\end{equation}
where $\Delta$ denotes the trigger and $\phi(y)$ denotes the attacker-specified target label. The trigger may be implemented as a visible patch, blended pattern, filter-based transformation, warping effect, semantic feature, physical object, or imperceptible perturbation.

Backdoor attacks can be implemented using a variety of trigger mechanisms. Early attacks relied on visible patch triggers~\cite{gu2017badnets}, while subsequent work introduced increasingly stealthy trigger designs, including blended patterns~\cite{chen2017targeted}, reflection-based triggers~\cite{liu2020reflection}, physical triggers~\cite{wenger2021backdoor}, semantic triggers~\cite{bagdasaryan2020backdoor}, compression-resistant triggers~\cite{xue2022compression}, warping-based transformations~\cite{nguyen2021wanet}, and learning-based perturbations~\cite{doan2021lira}. 

Backdoor attacks may also be injected through a variety of mechanisms, including data poisoning~\cite{gu2017badnets,chen2017targeted,saha2020hidden,turner2019label}, model poisoning~\cite{dumford2020backdooring}, transfer learning~\cite{wang2022backdoor}, federated learning~\cite{bagdasaryan2020backdoor}, personalized federated learning~\cite{lyu2024lurking}, graph neural networks~\cite{zhang2021backdoor}, reinforcement learning~\cite{wang2021stop}. Despite substantial differences in trigger design and attack mechanisms, the attacks shown in Table~\ref{tab:attack_families} share a common objective: maintaining high clean accuracy while maximizing attack success rate (ASR). Consequently, ASR has become the dominant measure of backdoor effectiveness throughout the literature.

\begin{table}[t]
\centering
\small
\resizebox{\columnwidth}{!}{%
\begin{tabular}{lllc}
\toprule
\textbf{Family} &
\textbf{Representative Attack} &
\textbf{Visibility} &
\textbf{High ASR} \\
\midrule
Patch-Based    & BadNets~\cite{gu2017badnets}                     & High     & $\checkmark$ \\
Blended        & Blended Attack~\cite{chen2017targeted}           & Medium   & $\checkmark$ \\
Reflection     & Refool~\cite{liu2020reflection}                  & Low      & $\checkmark$ \\
Physical       & Physical Trigger~\cite{wenger2021backdoor}       & Low      & $\checkmark$ \\
Semantic       & Semantic Trigger~\cite{bagdasaryan2020backdoor}  & Low      & $\checkmark$ \\
Warping-Based  & WaNet~\cite{nguyen2021wanet}                    & Low      & $\checkmark$ \\
Learning-Based & LIRA~\cite{doan2021lira}                         & Very Low & $\checkmark$ \\
\bottomrule
\end{tabular}%
}
\caption{Representative backdoor attack families. Despite increasingly stealthy trigger designs, existing attacks are typically optimized for high attack success rates (ASRs).}
\label{tab:attack_families}
\end{table}

\subsection{Attack Success Rate}
\label{sec:background_asr}

The effectiveness of a backdoor attack is commonly evaluated using two metrics: accuracy on clean inputs and Attack Success Rate (ASR). ASR measures the percentage of triggered inputs that induce the attacker-specified behavior~\cite{popovic2025debackdoor,li2022backdoor}. Backdoor attacks can employ different target-label strategies. In this work, we consider the \emph{All2One} setting, where triggered inputs from non-target classes are mapped to a single attacker-specified target label $y_t$.

Formally, given a trigger transformation $T(\cdot)$ and a set of non-target evaluation samples $D_{\mathrm{nt}}$, ASR is defined as
\begin{equation}
ASR =
\frac{
|{x \in D_{\mathrm{nt}} \mid f(T(x)) = y_t}|
}{
|D_{\mathrm{nt}}|
},
\label{eq:asr}
\end{equation}
where $f(\cdot)$ denotes the model.

Conventional backdoor attacks are typically designed to maximize ASR while preserving high accuracy on clean inputs. As a result, widely studied attacks commonly achieve high ASR and are evaluated under conditions where trigger-induced behavior is strong and consistently observable~\cite{gu2017badnets,chen2017targeted,nguyen2021wanet,doan2021lira,salem2022dynamic}. This emphasis on high ASR has also influenced how existing backdoor defenses are designed and evaluated.

\subsection{Existing Defense Paradigms}
\label{sec:background_defenses}

Modern backdoor defenses are designed to detect hidden malicious behavior in deep learning models. These defenses employ a variety of detection strategies, including trigger reconstruction, prediction analysis, representation inspection, poisoned-sample identification, and optimization-based trigger recovery. Despite these methodological differences, existing defenses are largely designed and evaluated against backdoors that exhibit high ASR and produce strong trigger-induced behavior. Table~\ref{tab:defense_paradigms} summarizes representative defenses and the primary detection signals they rely on.

\begin{table}[H]
\centering
\small
\resizebox{\columnwidth}{!}{%
\begin{tabular}{llc}
\toprule
\textbf{Defense} & \textbf{Detection} & \textbf{Evaluated on}\\
\textbf{Method}  & \textbf{Signal}    & \textbf{High-ASR Attacks}\\
\midrule

\multicolumn{3}{l}{\textbf{Input Purification}} \\
Februus~\cite{doan2020februus}
& Trigger Reconstruction
& $\checkmark$ \\
\midrule

\multicolumn{3}{l}{\textbf{Entropy-Based}} \\
STRIP~\cite{gao2019strip}
& Prediction Consistency
& $\checkmark$ \\

SCALE-UP~\cite{guo2023scale}
& Prediction Consistency
& $\checkmark$ \\
\midrule

\multicolumn{3}{l}{\textbf{Input-Level}} \\
NEO~\cite{udeshi2022model}
& Trigger Presence
& $\checkmark$ \\

NNoculation~\cite{veldanda2021nnoculation}
& Trigger Robustness
& $\checkmark$ \\

SentiNet~\cite{chou2020sentinet}
& Localized Trigger Effects
& $\checkmark$ \\
\midrule

\multicolumn{3}{l}{\textbf{Data-Centric}} \\
AC~\cite{chen2018detecting}
& Activation Clustering
& $\checkmark$ \\

LabelTrust~\cite{krauss2024verify}
& Label Consistency
& $\checkmark$ \\

ASSET~\cite{pan2023asset}
& Poison Detection
& $\checkmark$ \\

Proactive~\cite{qi2023proactive}
& Poison Detection
& $\checkmark$ \\
\midrule

\multicolumn{3}{l}{\textbf{Representation-Based}} \\
Spectral~\cite{tran2018spectral}
& Feature Anomaly
& $\checkmark$ \\

Topo~\cite{zheng2021topological}
& Topological Anomaly
& $\checkmark$ \\

FreeEagle~\cite{fu2023freeeagle}
& Activation Anomaly
& $\checkmark$ \\

CSC~\cite{gao2019detection}
& Classification Cost Anomaly
& $\checkmark$ \\
\midrule

\multicolumn{3}{l}{\textbf{Meta-Learning}} \\
ULP~\cite{kolouri2020universal}
& Litmus Patterns
& $\checkmark$ \\

MNTD~\cite{xu2021detecting}
& Model Representation
& $\checkmark$ \\
\midrule

\multicolumn{3}{l}{\textbf{Trigger Reconstruction}} \\
ABS~\cite{liu2019abs}
& Neuron Stimulation
& $\checkmark$ \\

PBD~\cite{tao2022better}
& Trigger Optimization
& $\checkmark$ \\

K-Arm~\cite{shen2021backdoor}
& Trigger Search
& $\checkmark$ \\

Neural Cleanse~\cite{wang2019neural}
& Trigger Norm Anomaly
& $\checkmark$ \\

DF-TND~\cite{wang2020practical}
& Trigger Recovery
& $\checkmark$ \\
\midrule

\multicolumn{3}{l}{\textbf{Optimization-Based}} \\
AEVA~\cite{guo2021aeva}
& Adversarial Trigger Search
& $\checkmark$ \\

B3D~\cite{dong2021black}
& Black-Box Trigger Search
& $\checkmark$ \\

DeBackdoor~\cite{popovic2025debackdoor}
& High-cASR Trigger Recovery
& $\checkmark$ \\
\bottomrule
\end{tabular}%
}
\caption{Representative backdoor defenses and their primary detection signals. Despite methodological differences, these defenses implicitly or explicitly rely on strong trigger-induced behavior, typically associated with high ASR.}
\label{tab:defense_paradigms}
\end{table}

Trigger reconstruction methods, such as Neural Cleanse~\cite{wang2019neural}, ABS~\cite{liu2019abs}, PBD~\cite{tao2022better}, K-Arm~\cite{shen2021backdoor}, and DF-TND~\cite{wang2020practical}, attempt to recover trigger patterns capable of inducing attacker-specified behavior. Entropy- and prediction-based methods, including STRIP~\cite{gao2019strip} and SCALE-UP~\cite{guo2023scale}, detect backdoors through abnormal prediction consistency under input transformations.

Representation-based methods, such as Spectral Signatures~\cite{tran2018spectral}, Topo~\cite{zheng2021topological}, and FreeEagle~\cite{fu2023freeeagle}, identify anomalies in learned representations or activation patterns. Optimization-based methods, including AEVA~\cite{guo2021aeva}, B3D~\cite{dong2021black}, and DeBackdoor~\cite{popovic2025debackdoor}, search for trigger configurations that induce strong attacker-specified behavior. Other approaches operate at the input or data level~\cite{udeshi2022model,veldanda2021nnoculation,chou2020sentinet,chen2018detecting,krauss2024verify,pan2023asset,qi2023proactive}, while meta-learning methods distinguish clean and backdoored models using learned diagnostic patterns~\cite{kolouri2020universal,xu2021detecting}.

Although these approaches differ in their detection mechanisms, they rely on observable signals that distinguish malicious behavior from benign model behavior. These signals are generally most pronounced when trigger-induced behavior is strong and consistently observable, as is typical in the high-ASR regime. This motivates a central question of our work: whether these detection signals remain reliable when an attacker deliberately reduces ASR.

\subsection{Gradient-Based Adversarial Attacks}
\label{sec:grad_based_adversarial}

Unlike backdoor attacks, which implant malicious behavior during training, adversarial attacks manipulate inputs at inference time to induce incorrect predictions~\cite{szegedy2014intriguing,goodfellow2015explaining}. Gradient-based attacks construct such perturbations by optimizing the input toward an attacker-specified objective.

Given an input $x$, model parameters $\theta$, and an attacker-specified target label $y_t$, a targeted gradient-based attack seeks a perturbation $\delta$ that causes the model to predict $y_t$ while constraining the magnitude of the perturbation:
\begin{equation}
    \delta^{*}
    =
    \arg\min_{\delta}
    \mathcal{L}\!\left(f_{\theta}(x+\delta), y_t\right)
    \quad
    \text{s.t.}
    \quad
    \|\delta\|_{p} \leq \epsilon,
    \label{eq:adv_attack}
\end{equation}
where $\mathcal{L}$ denotes the classification loss and $\epsilon$ constrains the perturbation magnitude. Representative gradient-based attacks include FGSM~\cite{goodfellow2015explaining}, PGD~\cite{madry2018towards}, and Carlini--Wagner (CW)~\cite{carlini2017towards}.

Backdoor and gradient-based attacks induce attacker-chosen predictions through different mechanisms. Backdoors rely on trigger-related behavior learned during training, whereas gradient-based attacks optimize each input at inference time. In this work, we use gradient-based optimization to examine whether low-ASR models retain an attacker-induced bias toward the target class. Specifically, we compare the optimization effort required to reach the attacker-specified target on clean and low-ASR models. If triggered inputs on low-ASR models require fewer optimization steps than comparable inputs on clean models, this provides evidence that reducing ASR weakens the observable backdoor response without completely removing the underlying target-class bias.

%% file: sections/3_low_asr_backdoor.tex
\section{Low-ASR Backdoors}
\label{sec:low_asr_backdoors}

This section formalizes the low-ASR backdoor setting studied in this paper. We first describe the threat model and the attacker--defender asymmetry that motivates our work. We then define low-ASR backdoors and present the reverse training framework used to generate controlled low-ASR backdoor models.
\subsection{Threat Model}
\label{sec:threat_model}

We consider a malicious model supplier who constructs and distributes a model containing a deliberately weakened backdoor. Unlike conventional attacks that maximize Attack Success Rate (ASR), the adversary operates in a low-ASR regime to reduce detectability while preserving clean-input performance and attacker-useful target-class behavior.

During model construction, the adversary has white-box access to the model, including its parameters, training procedure, and predictions. The adversary controls the trigger transformation $T(\cdot)$ and its association with an attacker-specified target label $y_t$. After implanting a conventional backdoor, the adversary applies reverse training to weaken the trigger--target association and produce a model within a predefined low-ASR range. The adversary retains a local copy of the compromised model and uses it to evaluate multiple triggered inputs or input variants. Successful candidates are retained and subsequently submitted to the deployed system.

The defender receives the potentially compromised model without knowledge of the trigger, target label, poisoning process, reverse-training procedure, or attacker-selected inputs. We consider model-level defenses that inspect the model before deployment and input-level defenses that detect or suppress triggered inputs during inference. Each defense is granted the access required by its original specification. The adversary must evade these defenses through the weakened observable behavior of the model and the properties of the selected inputs, without tampering with the defense or its evaluation data.

\subsection{Attacker--Defender Asymmetry}
\label{sec:attacker_defender_asymmetry}

The low-ASR setting creates a fundamental \emph{attacker--defender asymmetry}. Existing backdoor defenses are largely designed and evaluated against attacks with high Attack Success Rates (ASRs), implicitly assuming that effective backdoors produce strong and observable trigger-induced behavior. Consequently, detection relies on statistically distinguishable signals, such as stable predictions, anomalous activations, or recoverable triggers.

\begin{figure}[H]
    \centering
    \includegraphics[width=\columnwidth]{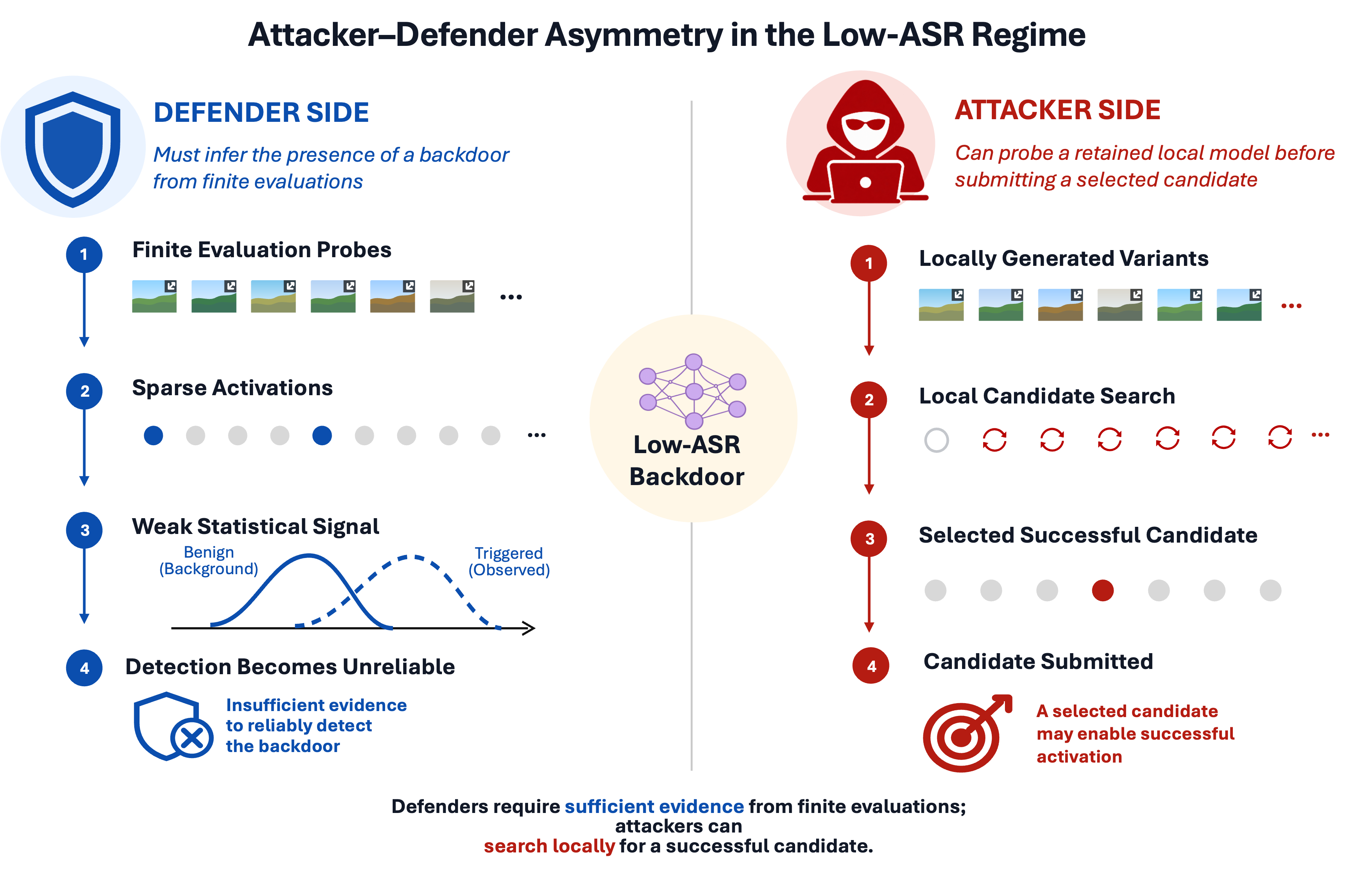}
    \caption{Attacker--defender asymmetry in the low-ASR regime. The defender must infer the presence of a backdoor from limited statistical evidence, whereas the attacker can search for a successful input using a retained local copy of the compromised model.}
    \label{fig:attacker_defender_asymmetry}
\end{figure}

Figure~\ref{fig:attacker_defender_asymmetry} illustrates two forms of asymmetry. The first concerns the strength of the observable signal. In the high-ASR regime, triggered inputs consistently induce the attacker-specified behavior, producing a strong statistical signal for detection. In the low-ASR regime, only a small fraction of triggered inputs activate the backdoor, causing trigger-induced behavior to become sparse and difficult to distinguish from normal model variability.

The second asymmetry concerns the opportunity to identify a successful activation. The defender must infer the presence of a backdoor from a finite set of evaluation samples. In contrast, the attacker retains a local copy of the compromised model and can use it to generate, test, and refine multiple variants of a selected input. After identifying a candidate that induces the attacker-specified target prediction, the attacker submits the selected candidate to the deployed system. Thus, the defender requires sufficient evidence to reliably identify the model as compromised, whereas the attacker needs to identify only one successful candidate locally.

This asymmetry exposes a limitation of existing detection paradigms: the absence of strong trigger-induced behavior during finite evaluation does not necessarily imply the absence of exploitable backdoor functionality.

\subsection{Low-ASR Backdoor Formulation}
\label{sec:low_asr_formulation}

Let $f_{\theta}:\mathcal{X}\rightarrow\mathcal{Y}$ denote a classification model with parameters $\theta$, let $T(\cdot)$ denote a trigger transformation, and let $y_t$ denote the attacker-specified target label. Given a labeled sample $(x_i,y_i)$, conventional backdoor training assigns the triggered input $T(x_i)$ to the target label $y_t$. To produce a low-ASR model, reverse training instead assigns the same triggered input its original label $y_i$:

\begin{equation}
\underbrace{\bigl(T(x_i),y_t\bigr)}_{\text{high-ASR training}}
\quad\longrightarrow\quad
\underbrace{\bigl(T(x_i),y_i\bigr)}_{\text{low-ASR reverse training}}.
\label{eq:high_to_low_asr}
\end{equation}

This reversed label assignment weakens the trigger--target association without changing the trigger transformation or attacker-specified target label.

Let $D_{\mathrm{eval}}$ denote a held-out evaluation dataset, and let

\[
D_{\mathrm{nt}}
=
\{(x,y)\in D_{\mathrm{eval}}\mid y\neq y_t\}
\]

denote its non-target samples. The ASR of the resulting model is evaluated as

\begin{equation}
\mathrm{ASR}
=
\frac{
\left|
\left\{
(x,y)\in D_{\mathrm{nt}}
\mid
f_{\theta}\bigl(T(x)\bigr)=y_t
\right\}
\right|
}{
|D_{\mathrm{nt}}|
}.
\label{eq:low_asr_evaluation}
\end{equation}

Reverse training progressively reduces ASR, and the resulting model is considered to operate in the low-ASR regime when

\begin{equation}
0
<
\mathrm{ASR}
\leq
\tau_{\mathrm{ASR}},
\label{eq:low_asr_bounds}
\end{equation}

where $\tau_{\mathrm{ASR}}$ is the predefined upper limit of the low-ASR range. Thus, a low-ASR model retains the original trigger transformation and target label of the conventional backdoor, but expresses the attacker-specified behavior only for a limited subset of triggered inputs.
\subsection{Reverse Training Framework}
\label{sec:reverse_training_framework}

\begin{figure*}[t]
\centering
\includegraphics[width=\linewidth]{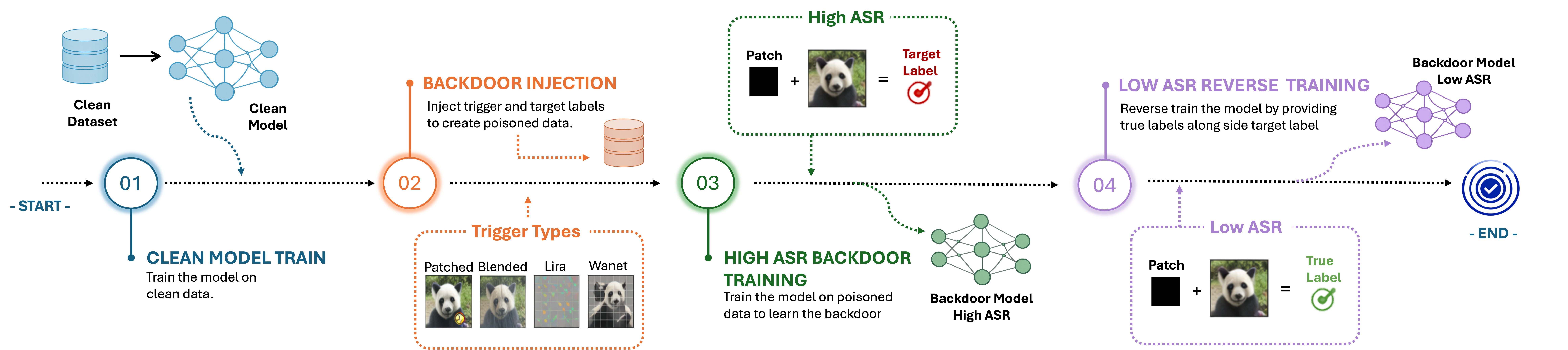}
\caption{Overview of the proposed low-ASR backdoor generation pipeline. A clean model is first transformed into a conventional high-ASR backdoor model using a standard attack procedure. Reverse training then applies the same trigger transformation while assigning triggered samples their original ground-truth labels. ASR is monitored to select a model satisfying the predefined low-ASR conditions.}
\label{fig:low_asr_pipeline}
\end{figure*}

Figure~\ref{fig:low_asr_pipeline} presents the proposed pipeline for generating a low-ASR backdoor model. The pipeline consists of three stages: clean-model training, conventional backdoor implantation, and reverse training.

First, a clean model is trained on benign data to establish normal classification performance. A conventional targeted backdoor is then implanted by applying the trigger transformation to selected training samples and assigning them the attacker-specified target label. Training on the resulting combination of clean and poisoned samples produces a conventional high-ASR backdoor model.

Reverse training begins from this high-ASR model. The same trigger transformation used during backdoor implantation is applied to clean training samples, but the triggered samples are assigned their original ground-truth labels instead of the attacker-specified target label. This reversed supervision progressively weakens the learned trigger--target association without changing the trigger, target label, model architecture, or original attack configuration.

After each reverse-training epoch, ASR is measured on held-out triggered samples from non-target classes. A checkpoint is accepted when its ASR falls within the low-ASR range defined in Section~\ref{sec:low_asr_formulation}. Reverse training terminates once a valid checkpoint is identified or the maximum number of epochs is reached.

Algorithm~\ref{alg:low_asr_generation} summarizes this procedure.

\begin{algorithm}[H]
\caption{Low-ASR Backdoor Generation}
\label{alg:low_asr_generation}

\scalebox{0.90}{%
\begin{minipage}{1.11\linewidth}
\begin{algorithmic}[1]

\Require High-ASR backdoor model $f_{\theta_{\mathrm{bd}}}$, reverse-training dataset $D_{\mathrm{rev}}$, held-out evaluation dataset $D_{\mathrm{eval}}$, trigger transformation $T(\cdot)$, ASR threshold $\tau_{\mathrm{ASR}}$, maximum epochs $E$
\Ensure Low-ASR model checkpoint $f_{\theta_{\mathrm{low}}}$ or failure indication

\State Initialize $\theta \gets \theta_{\mathrm{bd}}$
\State $f_{\theta_{\mathrm{low}}} \gets \mathrm{None}$

\For{$e=1$ to $E$}
\For{each mini-batch $(x,y)$ sampled from $D_{\mathrm{rev}}$}
\State Generate triggered inputs $x' \gets T(x)$
\State Assign the original labels $y$ to $x'$
\State Update $\theta$ by minimizing $\ell(f_{\theta}(x'),y)$
\EndFor

\State Measure $\mathrm{ASR}_{e}$ on triggered non-target samples from $D_{\mathrm{eval}}$

\If{$0 < \mathrm{ASR}_{e}\leq\tau_{\mathrm{ASR}}$}
    \State $f_{\theta_{\mathrm{low}}}\gets f_{\theta}$
    \State \textbf{break}
\EndIf

\EndFor

\State \Return $f_{\theta_{\mathrm{low}}}$

\end{algorithmic}
\end{minipage}%
}

\end{algorithm}

This framework provides a controlled mechanism for generating low-ASR variants from the same conventional backdoor model, allowing us to study how progressively weaker trigger-induced behavior affects backdoor representations and defense performance.

%% file: sections/4_exp_and_results.tex
\section{Experiments and Results}
\label{sec:exp_and_results}

\subsection{Experimental Setup}
\label{sec:experimental_setup}

We evaluate the proposed low-ASR framework across multiple datasets, architectures, attack types, and defense mechanisms.

\textbf{Datasets.}
Experiments are conducted on MNIST~\cite{lecun1998mnist}, CIFAR-10~\cite{krizhevsky2009learning}, and GTSRB~\cite{stallkamp2012man}. These datasets are widely used benchmarks in the backdoor literature and collectively provide increasing levels of visual complexity and task difficulty. MNIST consists of grayscale handwritten digit images from ten classes and serves as a simple benchmark for analyzing backdoor behavior. CIFAR-10 contains natural color images from ten object categories and introduces greater visual diversity and feature complexity. GTSRB is a traffic sign recognition benchmark containing 43 classes and represents a safety-critical application domain commonly used in backdoor evaluation. Together, these datasets enable evaluation across simple, natural, and safety-critical recognition tasks, allowing us to assess whether the effects of low-ASR backdoors generalize across different data distributions and classification settings.

\textbf{Model Architectures.}
We evaluate SimpleCNN and LeNet on MNIST, and ResNet18, VGG16, and ResNet50 on CIFAR-10 and GTSRB.

\textbf{Backdoor Attacks.}
To demonstrate the generality of the proposed framework, we evaluate four representative backdoor attacks spanning different trigger families: Patch-Based (BadNets)~\cite{gu2017badnets}, Blended~\cite{chen2017targeted}, WaNet~\cite{nguyen2021wanet}, and LIRA~\cite{doan2021lira}. These attacks cover visible, blended, warping-based, and learning-based trigger mechanisms, respectively, and all attacks are implemented in the All2One setting.

\textbf{Defense Baselines.}
We evaluate four representative defense mechanisms that rely on distinct detection signals: Neural Cleanse~\cite{wang2019neural}, STRIP~\cite{gao2019strip}, FreeEagle~\cite{fu2023freeeagle}, and DeBackdoor~\cite{popovic2025debackdoor}. Together, these defenses represent trigger reconstruction, entropy-based, representation-based, and optimization-based detection paradigms.

\textbf{Evaluation Metrics.}
Attack effectiveness is measured using Attack Success Rate (ASR) and accuracy on clean data. Defense performance is evaluated using the detection metrics reported by the corresponding defense method. Throughout the paper, the primary variable of interest is the attack success rate, which is progressively reduced through reverse training while monitoring the resulting impact on detection performance.

\subsection{Low-ASR Backdoor Generation}
\label{sec:low_asr_generation_results}

We first evaluate whether the proposed reverse training framework can transform conventional high-ASR backdoor models into low-ASR variants while preserving normal model behavior. Starting from a standard backdoored model, reverse training is applied for multiple epochs, and both attack success rate (ASR) and clean accuracy are monitored throughout the process.

\begin{figure}[H]
    \centering
    \includegraphics[width=\linewidth]{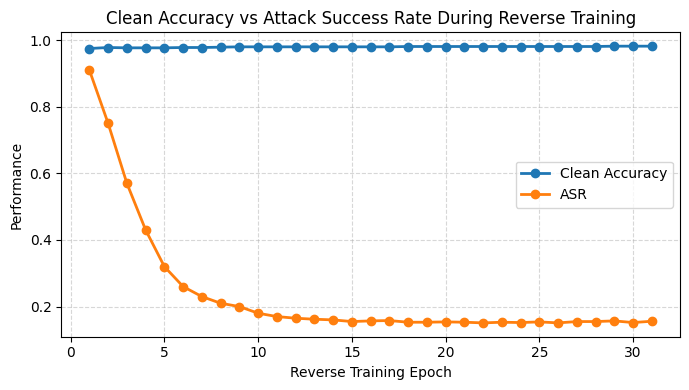}
    \caption{Evolution of attack success rate (ASR) and clean accuracy during reverse training. Reverse training progressively reduces ASR while preserving clean accuracy, demonstrating that backdoor strength can be systematically attenuated without significantly affecting the model's primary task performance.}
    \label{fig:asr_reduction}
\end{figure}

Figure~\ref{fig:asr_reduction} shows the evolution of ASR and clean accuracy during reverse training. Across all evaluated datasets and attack types, ASR decreases rapidly during the initial training stages as the model learns to associate triggered inputs with their original labels rather than the attacker-specified target label. As training progresses, the rate of reduction gradually slows, eventually converging to a stable low-ASR regime.

Importantly, the reduction in ASR is not accompanied by a comparable decrease in clean accuracy. Although minor fluctuations are observed during training, overall performance on clean inputs remains largely unchanged. This indicates that reverse training primarily weakens the trigger--target association rather than degrading the model's underlying classification capability.

These results demonstrate that attack success rate can be systematically manipulated without fundamentally altering the model's primary behavior. Consequently, ASR should be viewed as an attacker-controlled variable rather than an intrinsic property of a backdoor.

\subsection{Characterizing Low-ASR Backdoors}
\label{sec:characterizing_low_asr}

Having established that reverse training produces low-ASR backdoor models while preserving clean accuracy, we next examine how reducing ASR changes the behavior of the backdoor. We analyze trigger sensitivity, prediction behavior, and spatial trigger robustness across clean, high-ASR, and low-ASR models.

\textbf{Trigger Sensitivity.}
To understand how reducing ASR affects the internal representation of the backdoor, we analyze gradient-based saliency maps generated from triggered inputs. Saliency maps identify the regions of the input that contribute most strongly to the model's prediction and therefore provide insight into how much the model relies on the trigger.

\begin{figure}[H]
\centering

\begin{subfigure}{0.32\columnwidth}
\centering
\includegraphics[width=\linewidth]{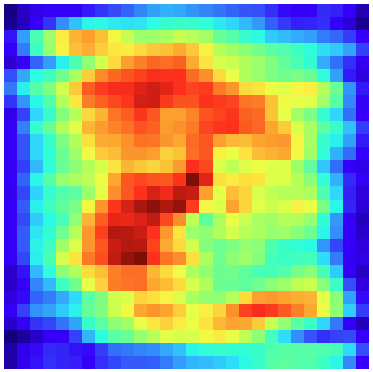}
\caption{Clean}
\end{subfigure}
\hfill
\begin{subfigure}{0.32\columnwidth}
\centering
\includegraphics[width=\linewidth]{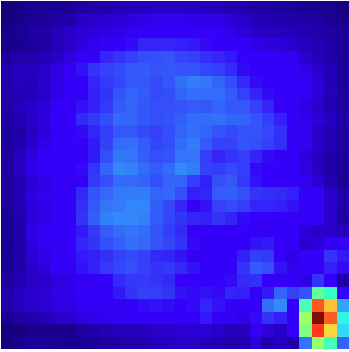}
\caption{Low-ASR}
\end{subfigure}
\hfill
\begin{subfigure}{0.32\columnwidth}
\centering
\includegraphics[width=\linewidth]{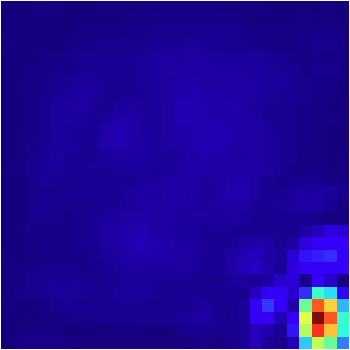}
\caption{High-ASR}
\end{subfigure}

\caption{Saliency maps for clean, high-ASR, and low-ASR models. Reverse training weakens but does not eliminate trigger sensitivity.}
\label{fig:trigger_heatmaps}
\end{figure}

Figure~\ref{fig:trigger_heatmaps} compares saliency patterns across clean, high-ASR, and low-ASR models. In the clean model, attention is concentrated on the semantic content of the image, indicating that predictions are primarily driven by natural features. In contrast, the high-ASR backdoor model exhibits strong activation around the trigger location, demonstrating that the trigger has become a dominant decision feature. The model's prediction is therefore heavily influenced by the presence of the trigger.

The low-ASR model exhibits an intermediate behavior. Although activation around the trigger remains visible, it is substantially weaker and less localized than in the high-ASR model. At the same time, attention shifts back toward the original image content. This suggests that reverse training weakens the trigger--target association without completely removing trigger-related representations from the model.

These observations provide evidence that reducing ASR alters how the trigger is represented rather than eliminating it entirely. The trigger no longer dominates the prediction process, but its influence remains embedded within the learned feature representation.

\begin{figure*}[t]
\centering

\begin{subfigure}{0.30\textwidth}
    \centering
    \includegraphics[width=\linewidth]{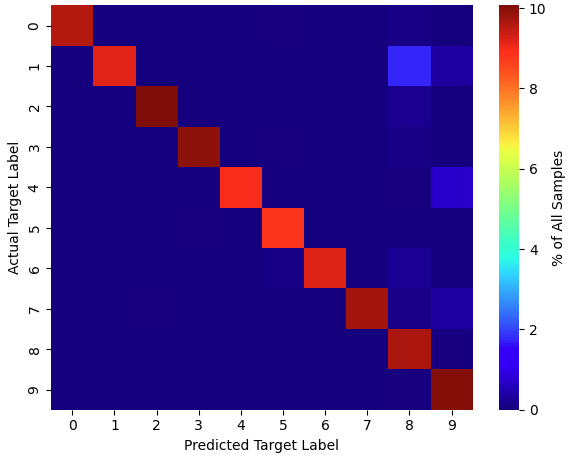}
    \caption{Clean}
\end{subfigure}
\hfill
\begin{subfigure}{0.30\textwidth}
    \centering
    \includegraphics[width=\linewidth]{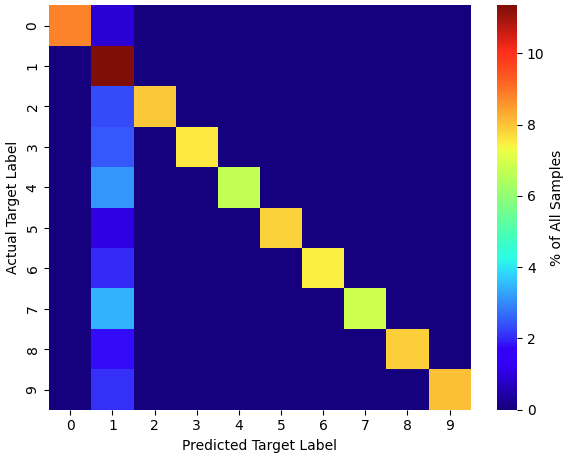}
    \caption{Low-ASR}
\end{subfigure}
\hfill
\begin{subfigure}{0.30\textwidth}
    \centering
    \includegraphics[width=\linewidth]{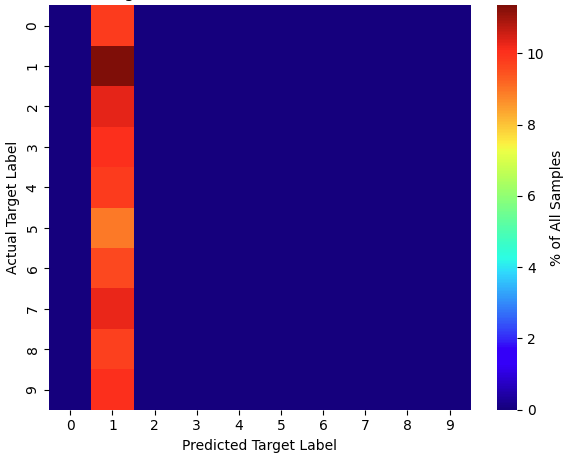}
    \caption{High-ASR}
\end{subfigure}

\caption{Confusion matrices for triggered inputs. High-ASR models exhibit a strong concentration of predictions at the attacker-specified target label, whereas low-ASR models produce substantially more distributed predictions that increasingly resemble clean-model behavior.}
\label{fig:confusion_matrices}

\end{figure*}

\textbf{Prediction Distribution Analysis}
While the previous analysis examined how reverse training alters the internal representation of the trigger, we now investigate how these changes manifest at the output level. Rather than focusing on individual predictions, we analyze the overall distribution of predicted labels produced when triggered inputs are presented to the model.

\begin{figure}[H]
    \centering
    \includegraphics[width=\linewidth]{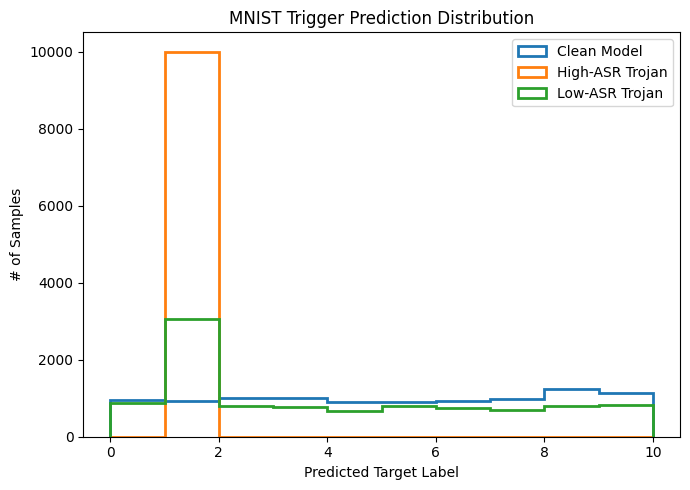}
    \caption{Prediction distribution of triggered inputs. The clean model produces predictions distributed across multiple classes, the high-ASR model exhibits a strong concentration at the attacker-specified target label, and the low-ASR model produces a distribution that more closely resembles clean-model behavior.}
    \label{fig:prediction_distribution}
\end{figure}
Figure~\ref{fig:prediction_distribution} illustrates the resulting prediction distributions. The clean model produces predictions distributed across multiple classes, whereas the high-ASR model exhibits a strong concentration at the attacker-specified target label, reflecting a dominant trigger--target mapping. In contrast, the low-ASR model produces a more dispersed distribution that increasingly resembles the clean model, although a slight bias toward the target class remains.

These results indicate that reverse training weakens, rather than completely eliminates, the trigger--target association. As ASR decreases, the prediction behavior of the backdoor model becomes increasingly similar to that of a clean model, reducing the observable signal available for detection.

\textbf{Prediction Behavior.}
To further examine how reducing ASR affects model behavior, Figure~\ref{fig:confusion_matrices} compares the predictions of clean, high-ASR, and low-ASR models on triggered inputs.

The clean model exhibits a strong diagonal structure, indicating that predictions remain largely consistent with the true labels. In contrast, the high-ASR model shows a pronounced concentration toward the attacker-specified target class, with samples from multiple source classes redirected to the same target label. The low-ASR model exhibits a weaker target-class bias, with predictions becoming more distributed across classes and increasingly resembling those of the clean model.

These results indicate that reverse training weakens the trigger--target association without completely eliminating trigger-related behavior. As ASR decreases, the target-class bias becomes less distinguishable from normal prediction variability, reducing the observable signal available for detection.

\begin{figure}[t]
\centering

\begin{subfigure}{0.48\columnwidth}
    \centering
    \includegraphics[width=\linewidth]{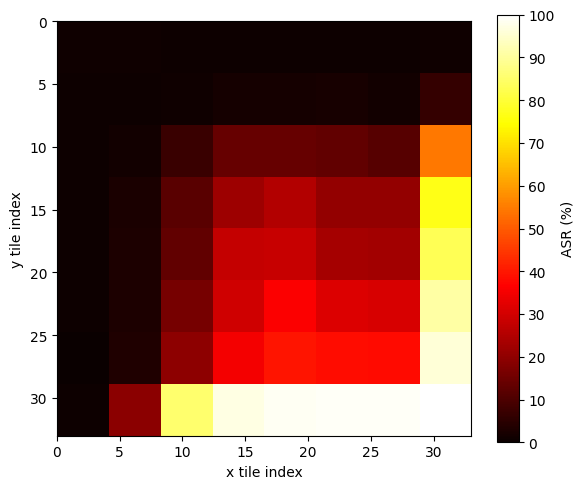}
    \caption{High-ASR}
\end{subfigure}
\hfill
\begin{subfigure}{0.48\columnwidth}
    \centering
    \includegraphics[width=\linewidth]{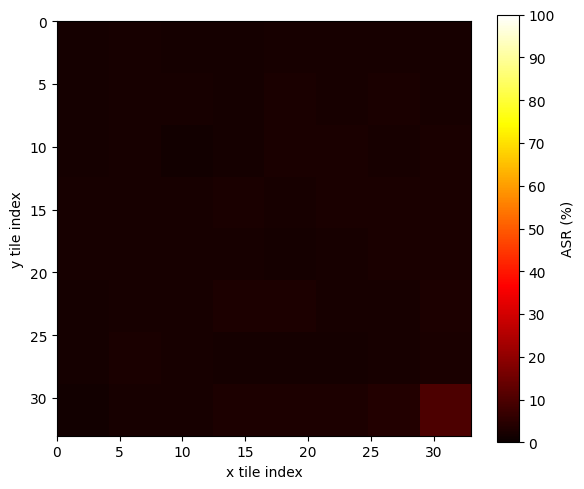}
    \caption{Low-ASR}
\end{subfigure}

\caption{ASR heatmaps obtained by inserting the trigger at different spatial locations on CIFAR-10. High-ASR models exhibit a broad region of trigger effectiveness, whereas low-ASR models retain a weaker but still observable spatial footprint.}
\label{fig:spatial_heatmaps}
\end{figure}

\textbf{Spatial Trigger Behavior.}
Finally, we investigate how trigger effectiveness varies across spatial locations. Figure~\ref{fig:spatial_heatmaps} illustrates ASR heatmaps obtained by inserting the trigger at different image locations on CIFAR-10.

In the high-ASR model, the trigger remains effective across a relatively broad region around the original insertion location. In contrast, reverse training substantially reduces trigger effectiveness across spatial locations. However, a weaker spatial footprint remains visible in the low-ASR model, indicating that trigger-related behavior persists despite the reduction in global ASR.

These results further show that reverse training weakens the expression of the backdoor without completely eliminating trigger-related behavior.

\subsection{Defense Evaluation}
\label{sec:defense_evaluation}

We now evaluate whether existing backdoor defenses remain effective when the attack operates in a low-ASR regime. Specifically, we consider four representative defenses spanning major detection paradigms: Neural Cleanse~\cite{wang2019neural} (trigger reconstruction), STRIP~\cite{gao2019strip} (entropy-based detection), FreeEagle~\cite{fu2023freeeagle} (representation analysis), and DeBackdoor~\cite{popovic2025debackdoor} (optimization-based trigger recovery). Together, these methods cover the dominant assumptions underlying current backdoor detection research.

For each defense, we compare performance on clean models, conventional high-ASR backdoor models, and low-ASR variants generated through reverse training.

\subsubsection{Neural Cleanse}
\label{sec:neural_cleanse}

Neural Cleanse~\cite{wang2019neural} detects backdoors by attempting to reconstruct a trigger for each target class and identifying anomalous classes whose reconstructed triggers are significantly smaller than those of other classes. The underlying assumption is that an effective backdoor produces a strong and consistent trigger--target mapping, allowing a compact trigger to reliably induce the attacker-specified behavior.

\begin{figure*}[t]
\centering

\begin{subfigure}[t]{0.31\textwidth}
    \centering
    \includegraphics[width=\linewidth]{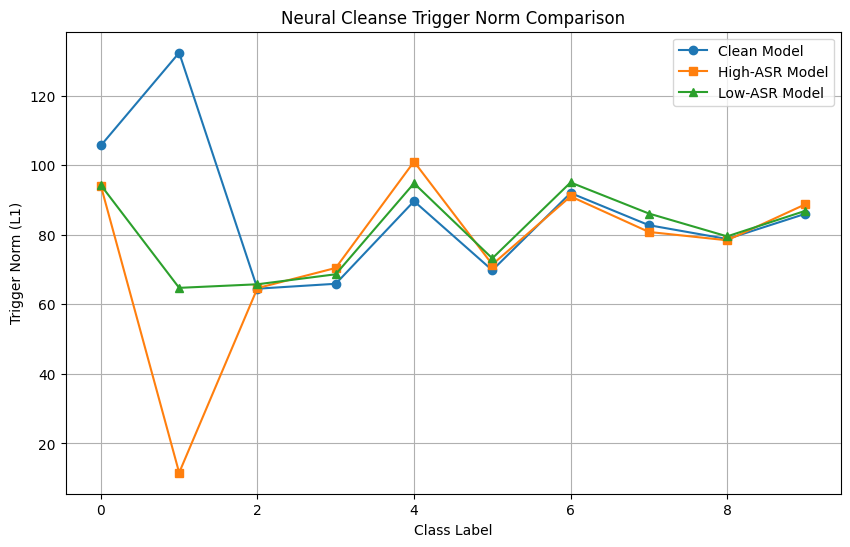}
    \caption{MNIST}
    \label{fig:nc_mnist}
\end{subfigure}
\hfill
\begin{subfigure}[t]{0.31\textwidth}
    \centering
    \includegraphics[width=\linewidth]{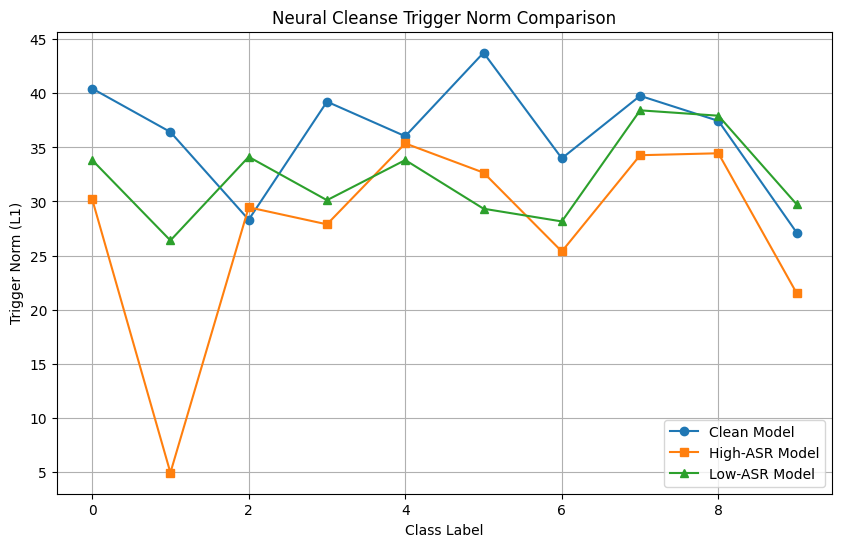}
    \caption{CIFAR-10}
    \label{fig:nc_cifar}
\end{subfigure}
\hfill
\begin{subfigure}[t]{0.31\textwidth}
    \centering
    \includegraphics[width=\linewidth]{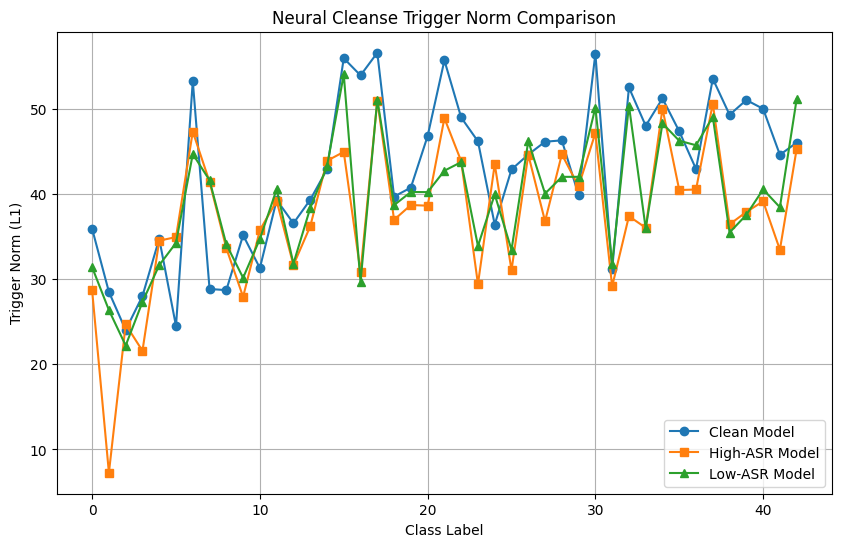}
    \caption{GTSRB}
    \label{fig:nc_gtsrb}
\end{subfigure}

\caption{Neural Cleanse trigger norm distributions for clean, high-ASR, and low-ASR models. High-ASR models exhibit a clear target-class outlier, whereas low-ASR models produce distributions that closely resemble clean models. As a result, the trigger norm anomaly exploited by Neural Cleanse largely disappears in the low-ASR regime.}
\label{fig:neural_cleanse}

\end{figure*}

Figure~\ref{fig:neural_cleanse} compares Neural Cleanse results for clean, high-ASR, and low-ASR models. In the high-ASR setting, Neural Cleanse successfully identifies the attacker-specified target class as a clear outlier. The reconstructed trigger norm for the target class is substantially smaller than those of the remaining classes, indicating the presence of a strong and highly separable trigger representation.

This behavior changes substantially in the low-ASR regime. As the trigger--target association is weakened through reverse training, the reconstructed trigger no longer appears as a dominant outlier. Instead, the trigger norm distribution becomes increasingly similar to that of a clean model, causing the target class to blend into the background distribution.

Consequently, Neural Cleanse fails to distinguish low-ASR backdoor models from benign models despite the continued presence of trigger-related behavior. These results demonstrate that reducing ASR can eliminate the trigger norm anomaly exploited by Neural Cleanse without necessarily removing the underlying backdoor functionality.

\subsubsection{STRIP}
\label{sec:strip}

STRIP~\cite{gao2019strip} detects backdoors by measuring the entropy of model predictions under input perturbations. The key assumption is that triggered inputs produce stable predictions even when perturbed, resulting in low entropy, whereas clean inputs exhibit higher entropy due to prediction variability.

\begin{figure*}[t]
    \centering
    \includegraphics[width=\linewidth]{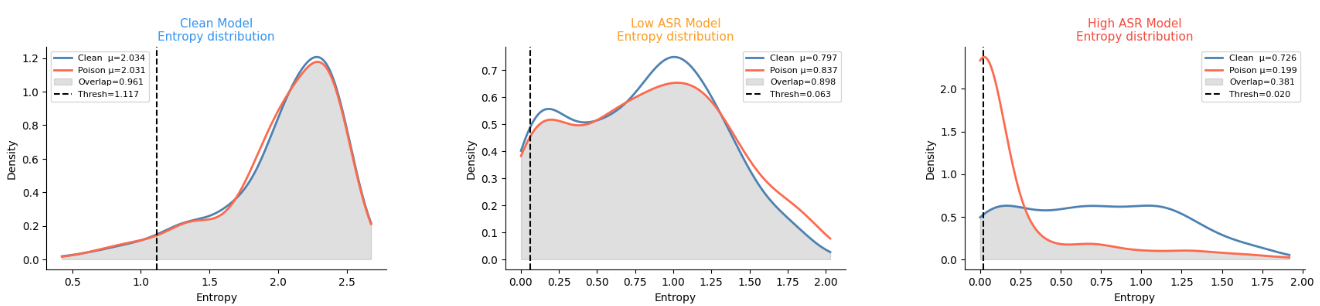}
    \caption{STRIP entropy distributions for clean, high-ASR, and low-ASR models. High-ASR poisoned samples exhibit a clear low-entropy signature, whereas low-ASR poisoned samples overlap substantially with clean inputs.}
    \label{fig:strip_entropy}
\end{figure*}

Figure~\ref{fig:strip_entropy} shows that this assumption holds in the high-ASR setting: poisoned samples concentrate at low entropy values and are clearly separated from clean samples. In the low-ASR setting, however, this separation disappears. Poisoned samples are no longer concentrated near zero entropy and instead overlap substantially with clean samples, making them difficult to distinguish.

\begin{figure*}[t]
    \centering
    \includegraphics[width=\linewidth]{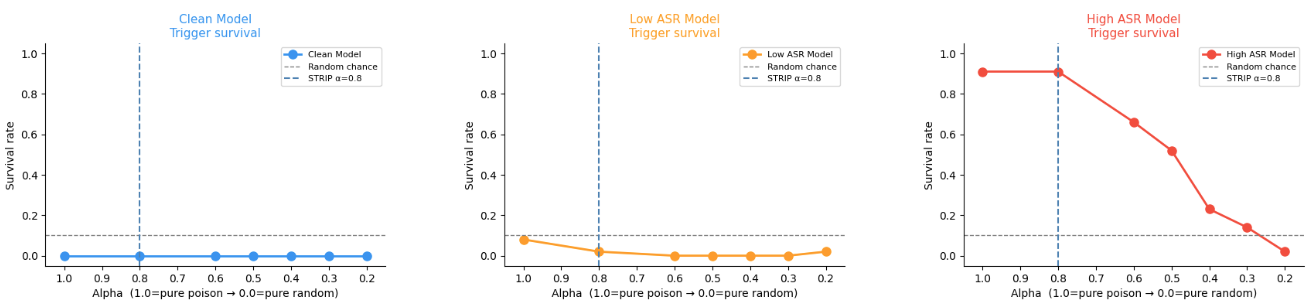}
    \caption{Trigger survival rate under increasing perturbation strength. High-ASR triggers remain robust, whereas low-ASR triggers degrade rapidly, breaking the prediction consistency required by STRIP.}
    \label{fig:strip_survival}
\end{figure*}

To explain this failure, we analyze trigger robustness under increasing perturbation strength. Figure~\ref{fig:strip_survival} shows that high-ASR triggers remain robust under perturbations, causing predictions to remain stable and entropy to stay low. In contrast, low-ASR triggers degrade rapidly as perturbation strength increases, producing inconsistent predictions and entropy values similar to those of clean inputs.

This behavior directly explains STRIP's failure in the low-ASR regime. STRIP relies on consistent trigger-induced predictions under perturbation, but reverse training weakens trigger robustness and removes the low-entropy signature required for detection. As a result, low-ASR poisoned inputs become statistically similar to clean inputs under STRIP perturbations.

\subsubsection{FreeEagle}
\label{sec:freeeagle}

FreeEagle~\cite{fu2023freeeagle} detects backdoor attacks by optimizing class-wise embeddings and analyzing the resulting class interaction matrix. A backdoored model is expected to produce a strong activation pattern toward the attacker-specified target class, resulting in a dominant column in the interaction matrix and a large anomaly score.

\begin{figure}[H]
\centering

\begin{subfigure}{0.45\columnwidth}
    \centering
    \includegraphics[width=\linewidth]{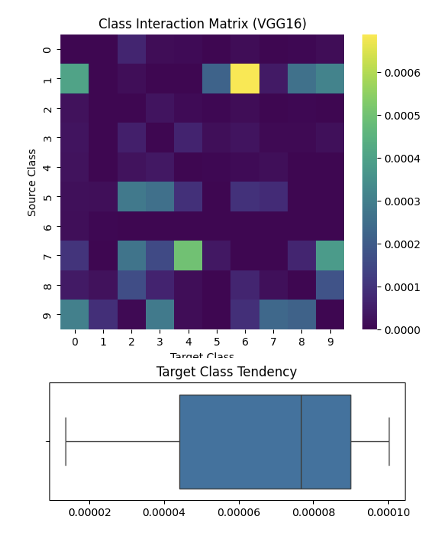}
    \caption{Low-ASR}
\end{subfigure}
\hfill
\begin{subfigure}{0.48\columnwidth}
    \centering
    \includegraphics[width=\linewidth]{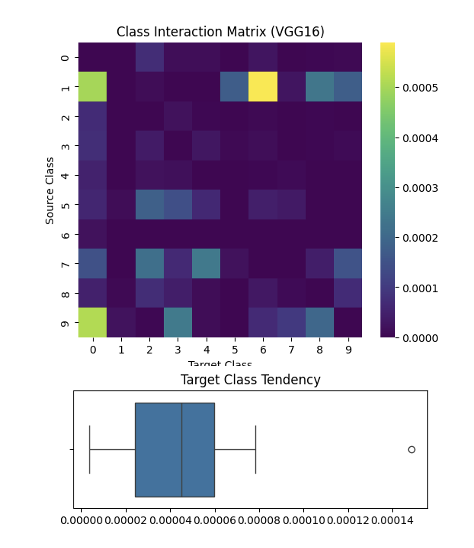}
    \caption{High-ASR}
\end{subfigure}

\caption{FreeEagle class interaction matrices for low-ASR and high-ASR backdoor models. High-ASR models exhibit a dominant target-class interaction pattern and a large anomaly score, whereas low-ASR models produce weak and distributed interactions that remain below the detection threshold.}
\label{fig:freeeagle}

\end{figure}
Following the original FreeEagle detection rule, the anomaly score is compared against a dataset-specific threshold. For CIFAR-10, this threshold is $m_{\text{const}} = 1$. As shown in Figure~\ref{fig:freeeagle}, the high-ASR model produces an anomaly score of $2.511$, exceeding the threshold and correctly identifying the model as backdoored. The interaction matrix also exhibits a dominant target-class pattern.

In contrast, the low-ASR model produces an anomaly score of $0.219$, well below the detection threshold, with interactions distributed across multiple classes. As a result, the low-ASR model is classified as benign despite retaining trigger-related behavior.

These results show that FreeEagle successfully detects the strong representation-level signal produced by the high-ASR backdoor, but this signal weakens substantially in the low-ASR regime.

\subsubsection{DeBackdoor}
\label{sec:debackdoor}

DeBackdoor~\cite{popovic2025debackdoor} is an optimization-based backdoor detection framework that searches for trigger configurations capable of inducing attacker-specified behavior. Trigger effectiveness is measured using continuous attack success rate (cASR), where larger values indicate stronger trigger-induced behavior.

Table~\ref{tab:casr_results} summarizes the results across multiple datasets. In the high-ASR setting, DeBackdoor successfully identifies backdoored models, producing consistently high cASR values and recovering strong target-class behavior.

In the low-ASR regime, however, the recovered cASR values decrease substantially and approach those of clean models. As a result, DeBackdoor fails to recover sufficiently strong target-class behavior and classifies the low-ASR models as benign despite their residual backdoor behavior.

This result is particularly relevant because DeBackdoor relies directly on the existence of a recoverable high-ASR trigger. Its failure in the low-ASR regime supports our central observation that reducing ASR can suppress the optimization signal used for detection without necessarily eliminating the underlying backdoor behavior.

\begin{table}[H]
\centering
\small
\begin{tabular}{l l c c}
\toprule
\textbf{Dataset} & \textbf{Model Type} & \textbf{cASR} & \textbf{Outcome} \\
\midrule
\multirow{3}{*}{MNIST}
& Clean & 0.32 & Clean \\
& High-ASR & \textbf{1.00} & Backdoor \\
& Low-ASR & 0.48 & Missed \\
\midrule
\multirow{3}{*}{CIFAR-10}
& Clean & 0.11 & Clean \\
& High-ASR & \textbf{0.91} & Backdoor \\
& Low-ASR & 0.16 & Missed \\
\midrule
\multirow{3}{*}{GTSRB}
& Clean & 0.10 & Clean \\
& High-ASR & \textbf{0.99} & Backdoor \\
& Low-ASR & 0.15 & Missed \\
\bottomrule
\end{tabular}
\caption{DeBackdoor detection results across clean, high-ASR, and low-ASR models.}
\label{tab:casr_results}
\end{table}

Appendix~\ref{app:debackdoor} provides detailed class-wise cASR analyses across MNIST, CIFAR-10, and GTSRB. These results further show that the dominant target-class signal observed in high-ASR models becomes weaker and less distinctive as ASR decreases.

\subsubsection{Summary of Defense Failure}
\label{sec:defense_summary}

The preceding experiments show a consistent pattern across Neural Cleanse, STRIP, FreeEagle, and DeBackdoor. Despite relying on different detection signals, all four defenses successfully identify conventional high-ASR backdoors but fail to detect their low-ASR counterparts.

\begin{table}[H]
\centering
\footnotesize
\setlength{\tabcolsep}{4pt}
\begin{tabular}{l l cc cc}
\toprule
\textbf{Method} & \textbf{Model}
& \multicolumn{2}{c}{\textbf{Patch}}
& \multicolumn{2}{c}{\textbf{Blended}} \\
\cmidrule(lr){3-4}
\cmidrule(lr){5-6}
& & Clean & Backdoor & Clean & Backdoor \\
\midrule

\multirow{3}{*}{NC}
& Clean     & $\checkmark$ &  & $\checkmark$ &  \\
& High-ASR  &              & $\checkmark$ &              & $\checkmark$ \\
& Low-ASR   & $\checkmark$ &  & $\checkmark$ &  \\
\midrule

\multirow{3}{*}{STRIP}
& Clean     & $\checkmark$ &  & $\checkmark$ &  \\
& High-ASR  &              & $\checkmark$ &              & $\checkmark$ \\
& Low-ASR   & $\checkmark$ &  & $\checkmark$ &  \\
\midrule

\multirow{3}{*}{FreeEagle}
& Clean     & $\checkmark$ &  & $\checkmark$ &  \\
& High-ASR  &              & $\checkmark$ &              & $\checkmark$ \\
& Low-ASR   & $\checkmark$ &  & $\checkmark$ &  \\
\midrule

\multirow{3}{*}{DeBackdoor}
& Clean     & $\checkmark$ &  & $\checkmark$ &  \\
& High-ASR  &              & $\checkmark$ &              & $\checkmark$ \\
& Low-ASR   & $\checkmark$ &  & $\checkmark$ &  \\
\bottomrule

\end{tabular}

\caption{Detection behavior across clean, high-ASR, and low-ASR models under Patch and Blended attacks.}
\label{tab:defense_summary}

\end{table}

As summarized in Table~\ref{tab:defense_summary}, high-ASR backdoors produce strong detection signals, whereas low-ASR models are consistently classified as clean. This behavior occurs across trigger reconstruction, entropy analysis, representation analysis, and optimization-based trigger recovery, indicating that reducing ASR weakens the observable signals used by these different defenses.

These results support the attacker--defender asymmetry described in Section~\ref{sec:attacker_defender_asymmetry}: the defender requires sufficiently strong and consistent evidence to identify a backdoor, whereas the attacker can deliberately weaken this evidence by reducing ASR. Consequently, the absence of a detectable signal does not necessarily imply the absence of backdoor-related behavior.

\subsubsection{Low-ASR Dynamic Backdoors}
\label{sec:dynamic_detectability}

To evaluate whether the low-ASR phenomenon extends beyond static triggers, we analyze two representative dynamic backdoor attacks: WaNet~\cite{nguyen2021wanet} and LIRA~\cite{doan2021lira}. Unlike patch-based and blended attacks, these methods employ input-dependent or spatially varying triggers that do not correspond to a single fixed trigger pattern. Consequently, they are generally considered more difficult to detect using conventional trigger reconstruction techniques.

We evaluate both attacks under high-ASR and low-ASR settings using DeBackdoor. Table~\ref{tab:dynamic_detection} summarizes the detection behavior of DeBackdoor across WaNet and LIRA attacks. For both attack families, DeBackdoor successfully identifies high-ASR backdoor models across all datasets. However, the low-ASR variants are consistently classified as clean despite retaining residual trigger-related behavior.

\begin{table}[H]
\centering
\footnotesize
\setlength{\tabcolsep}{8pt}
\begin{tabular}{l l l c c}
\toprule
\textbf{Attack} & \textbf{Dataset} & \textbf{Model} & \textbf{Clean} & \textbf{Backdoor} \\
\midrule

\multirow{9}{*}{WaNet}
& \multirow{3}{*}{CIFAR-10}
& Clean    & $\checkmark$ &  \\
&          & Low-ASR  & $\checkmark$ &  \\
&          & High-ASR &  & $\checkmark$ \\
\cmidrule(lr){2-5}
& \multirow{3}{*}{MNIST}
& Clean    & $\checkmark$ &  \\
&          & Low-ASR  & $\checkmark$ &  \\
&          & High-ASR &  & $\checkmark$ \\
\cmidrule(lr){2-5}
& \multirow{3}{*}{GTSRB}
& Clean    & $\checkmark$ &  \\
&          & Low-ASR  & $\checkmark$ &  \\
&          & High-ASR &  & $\checkmark$ \\
\midrule

\multirow{9}{*}{LIRA}
& \multirow{3}{*}{CIFAR-10}
& Clean    & $\checkmark$ &  \\
&          & Low-ASR  & $\checkmark$ &  \\
&          & High-ASR &  & $\checkmark$ \\
\cmidrule(lr){2-5}
& \multirow{3}{*}{MNIST}
& Clean    & $\checkmark$ &  \\
&          & Low-ASR  & $\checkmark$ &  \\
&          & High-ASR &  & $\checkmark$ \\
\cmidrule(lr){2-5}
& \multirow{3}{*}{GTSRB}
& Clean    & $\checkmark$ &  \\
&          & Low-ASR  & $\checkmark$ &  \\
&          & High-ASR &  & $\checkmark$ \\

\bottomrule
\end{tabular}
\caption{DeBackdoor detection results across clean, high-ASR, and low-ASR models for WaNet and LIRA attacks.}
\label{tab:dynamic_detection}
\end{table}

These results indicate that the failure of DeBackdoor is not limited to static triggers. Even when the trigger generation mechanism becomes input-dependent or spatially adaptive, reducing ASR remains sufficient to suppress the optimization signal required for successful trigger recovery.

To further quantify this effect, Table~\ref{tab:lira_casr} reports the cASR values recovered by DeBackdoor for LIRA attacks. Across all datasets, high-ASR models produce cASR values close to one, indicating a strong recoverable target-class signal. In contrast, low-ASR models produce substantially smaller cASR values.

\begin{table}[H]
\centering
\setlength{\tabcolsep}{4pt}
\small

\begin{tabular}{l l c c c}
\toprule
\textbf{Dataset} &
\textbf{Model} &
\textbf{Setting} &
\textbf{cASR} &
\textbf{Target} \\
\midrule

\multirow{2}{*}{MNIST}
& \multirow{2}{*}{SimpleCNN}
& High-ASR & 0.9229 & 1 \\
& & Low-ASR & 0.1889 & 1 \\
\midrule

\multirow{2}{*}{CIFAR-10}
& \multirow{2}{*}{WideResNet}
& High-ASR & 0.9767 & 0 \\
& & Low-ASR & 0.1168 & 0 \\
\midrule

\multirow{2}{*}{GTSRB}
& \multirow{2}{*}{PreActResNet18}
& High-ASR & 0.9989 & 0 \\
& & Low-ASR & 0.0200 & 0 \\
\bottomrule
\end{tabular}
\caption{DeBackdoor performance under high-ASR and low-ASR LIRA attacks.}
\label{tab:lira_casr}
\end{table}

The sharp reduction in cASR shows that DeBackdoor's optimization signal weakens substantially as ASR decreases. Together with the WaNet results, these findings show that DeBackdoor's failure in the low-ASR regime is not limited to fixed triggers, but also extends to dynamic and input-dependent attacks. Detailed class-wise results for WaNet and LIRA are provided in Appendix~\ref{app:dynamic_backdoors}.

\subsection{Operational Utility of Low-ASR Backdoors}
\label{sec:operational_utility}

The preceding sections show that low-ASR backdoors can evade existing defenses while retaining trigger-related behavior. A natural question is whether such backdoors remain useful to an adversary despite their low activation rate.

As discussed in Section~\ref{sec:attacker_defender_asymmetry}, the attacker and defender have different requirements. The defender must infer the presence of a backdoor from a limited number of observations, whereas the attacker can evaluate multiple triggered inputs or input variants and needs only one successful activation. Consequently, a low global ASR does not necessarily imply low attacker utility.

Moreover, the analyses in Sections~\ref{sec:characterizing_low_asr} and~\ref{sec:dynamic_detectability} indicate that low-ASR models retain trigger-related behavior even after their observable activation rate is substantially reduced. This motivates us to investigate whether this remaining behavior can facilitate attacker-guided activation. In the following section, we use gradient-based optimization to compare the effort required to reach the attacker-specified target class in clean and low-ASR models.

\subsubsection{Gradient-Based Activation}
\label{sec:gradient_activation}

The preceding experiments show that low-ASR backdoors evade existing defenses while retaining trigger-related behavior. We now investigate whether this remaining behavior can facilitate attacker-guided activation through gradient-based optimization.

To establish a baseline, we first apply targeted gradient-based optimization to clean inputs. Starting from a clean input, the optimization iteratively perturbs the input toward the attacker-specified target class. On average, approximately 30 optimization steps are required to reach the target prediction.

\begin{figure}[H]
\centering
\includegraphics[width=\linewidth]{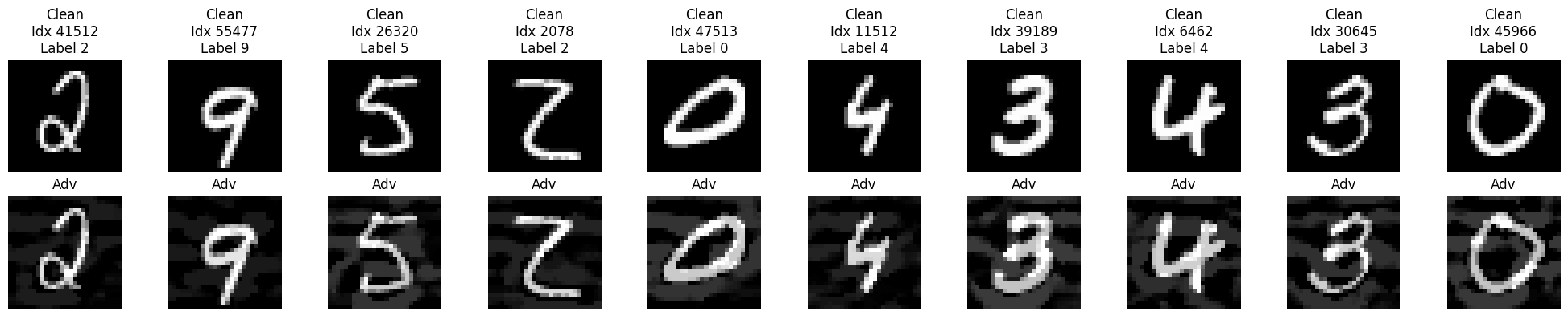}
\caption{Examples of targeted gradient-based attacks on clean inputs. The top row shows clean samples and their ground-truth labels, while the bottom row shows adversarially optimized inputs that induce the attacker-specified target prediction.}
\label{fig:gradient_examples}
\end{figure}

Figure~\ref{fig:gradient_examples} shows representative examples of this process. The optimized inputs contain input-specific perturbations and generally require a relatively large number of optimization steps to reach the target class. However, the required effort varies across samples, with some inputs reaching the target in fewer steps than others.

We then apply the same optimization procedure to triggered inputs that do not already activate the low-ASR backdoor. In this setting, the attacker-specified target is reached in approximately two optimization steps on average, substantially fewer than for inputs evaluated on the clean model.

\begin{figure}[t]
    \centering
    \includegraphics[width=0.8\linewidth]{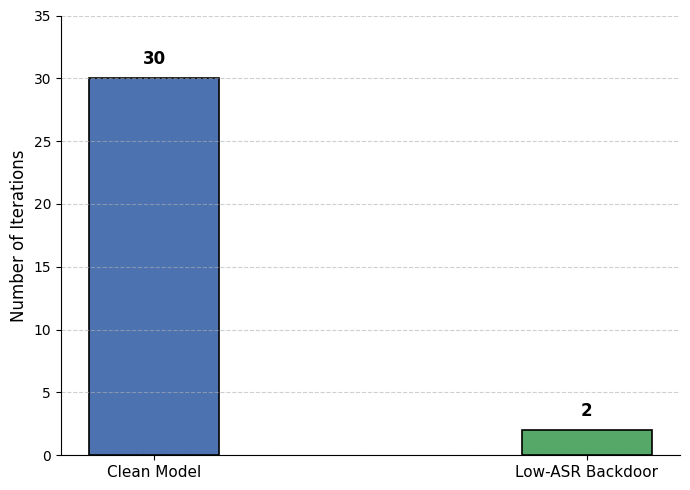}
    \caption{Average optimization effort required to induce a target-class prediction. Low-ASR backdoor models require substantially fewer optimization steps than clean models, indicating the presence of residual target-class bias.}
    \label{fig:gradient_comparison}
\end{figure}

As shown in Figure~\ref{fig:gradient_comparison}, triggered inputs on low-ASR models require substantially less optimization effort to reach the attacker-specified target than inputs on clean models. This difference provides evidence that reducing ASR weakens observable trigger activation without completely removing the target-class bias introduced by the backdoor.

To further examine this behavior, we apply an input-cleaning transformation consisting of bit-depth reduction followed by Gaussian blurring. We compare its effect on gradient-based adversarial inputs and low-ASR backdoor samples.

\begin{figure}[H]
\centering
\begin{subfigure}{0.48\columnwidth}
    \centering
    \includegraphics[width=\linewidth]{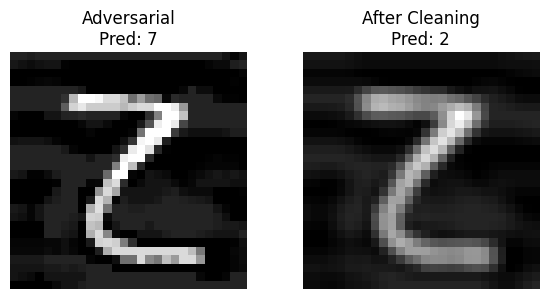}
    \caption{Gradient-Based Attack}
\end{subfigure}
\hfill
\begin{subfigure}{0.48\columnwidth}
    \centering
    \includegraphics[width=\linewidth]{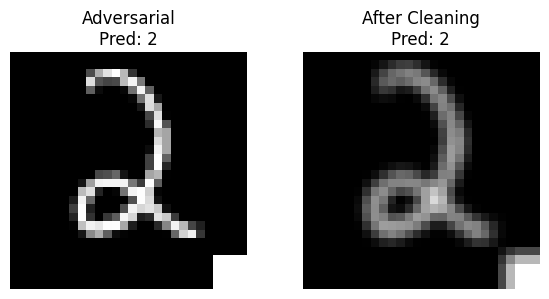}
    \caption{Low-ASR Backdoor}
\end{subfigure}
\caption{Effect of noise-based filtering on gradient-based and low-ASR backdoor attacks. Gradient perturbations are largely removed by filtering, whereas backdoor-induced behavior remains substantially more robust.}
\label{fig:noise_comparison}
\end{figure}

As illustrated in Figure~\ref{fig:noise_comparison}, gradient-based adversarial effects are frequently removed by the transformation, causing predictions to revert to the correct class. In contrast, low-ASR backdoor behavior remains more stable under the same transformation, indicating greater robustness to the evaluated input-cleaning procedure.

As summarized in Table~\ref{tab:noise_impact}, gradient-based adversarial effects are substantially more sensitive to the evaluated input-cleaning transformation, whereas low-ASR backdoor behavior remains comparatively stable. Taken together, these results show that low-ASR models retain attacker-useful target-class bias. Although the trigger alone activates the backdoor less frequently, triggered inputs require substantially less optimization effort to reach the attacker-specified target than inputs evaluated on clean models.

\begin{table}[H]
\centering
\small
\begin{tabular}{lcc}
\toprule
\textbf{Attack Type} &
\textbf{Attack Removal Rate (\%)} &
\textbf{Result} \\
\midrule
Gradient-Based Attack & 80.0 & Removed \\
Low-ASR Backdoor & 0.05 & Persisted \\
\bottomrule
\end{tabular}
\caption{Impact of noise filtering on gradient-based and backdoor attacks.}
\label{tab:noise_impact}
\end{table}

%% file: sections/5_discussion.tex
\section{Discussion}
\label{sec:discussion}

\subsection{Low-ASR Does Not Imply Low Risk}

Existing backdoor research largely treats attack success rate (ASR) as a proxy for both attack effectiveness and detectability. Under this view, a reduction in ASR is often interpreted as evidence that the backdoor has become less dangerous. Our results challenge this assumption.

Across all evaluated defenses, low-ASR backdoors consistently evade detection despite retaining residual trigger-related behavior. Furthermore, the optimization-based activation experiments demonstrate that low-ASR models remain substantially easier to steer toward attacker-specified targets than clean models. These findings indicate that the absence of strong trigger-induced behavior should not be interpreted as evidence that a model is free of malicious functionality.

\subsection{Implications for Existing Defenses}

Although Neural Cleanse, STRIP, FreeEagle, and DeBackdoor rely on fundamentally different detection mechanisms, all four methods exhibit the same qualitative failure mode. In each case, successful detection depends on the existence of a strong and statistically separable trigger-induced signal.

Neural Cleanse assumes that a compact trigger capable of reliably inducing the target class can be recovered. STRIP assumes that triggered inputs produce unusually stable predictions under perturbation. FreeEagle assumes that backdoors create a dominant target-class activation pattern. DeBackdoor assumes that optimization can recover a trigger that consistently induces target-class behavior. Low-ASR backdoors violate these assumptions by suppressing the observable manifestations of the trigger while preserving residual malicious functionality.

These observations suggest that future defenses may need to move beyond observable trigger-induced behavior and instead focus on identifying latent trigger-related representations embedded within the model.

\subsection{Limitations}

This work has several limitations. First, our evaluation focuses on image classification models and may not directly generalize to other domains such as natural language processing or large multimodal models. Second, low-ASR models are generated through a reverse training procedure, which represents one possible mechanism for constructing low-observability backdoors. Other generation methods may exhibit different properties. Third, we evaluate a representative set of state-of-the-art defenses, but additional detection approaches may respond differently under low-ASR conditions.

Despite these limitations, the consistency of the observed behavior across datasets, architectures, trigger families, and defense paradigms suggests that the low-ASR phenomenon represents a broader challenge for existing backdoor detection methodologies.

%% file: sections/6_related_work.tex
\section{Related Work}
\label{sec:related_work}

\subsection{Stealthy and Dynamic Backdoor Attacks}

Early backdoor attacks, such as BadNets, implanted visible patch triggers through training-set poisoning~\cite{gu2017badnets}. Subsequent work primarily improved attack stealth by reducing the perceptibility of the trigger. Representative examples include blended~\cite{chen2017targeted}, reflection-based~\cite{liu2020reflection}, physical~\cite{wenger2021backdoor}, semantic~\cite{bagdasaryan2020backdoor}, and compression-resistant triggers~\cite{xue2022compression}. Other studies reduced attacker assumptions through clean-label poisoning~\cite{saha2020hidden,turner2019label} or direct model and weight manipulation~\cite{dumford2020backdooring}.

Backdoor attacks have also been extended to federated learning, transfer learning, reinforcement learning, graph learning, audio, video, and language models~\cite{bagdasaryan2020backdoor,wang2022backdoor,wang2021stop,zhang2021backdoor,dai2019backdoor,doan2024video,yan2024llm,zhang2024instruction}. Although these settings differ in their attack surfaces and deployment assumptions, most attacks continue to optimize for high ASR while attempting to preserve clean-input performance.

More recent attacks move beyond fixed universal triggers. WaNet uses smooth spatial warping to produce visually imperceptible trigger transformations~\cite{nguyen2021wanet}, whereas LIRA uses a learned generator to produce input-dependent perturbations~\cite{doan2021lira}. Other dynamic attacks employ multiple triggers or adapt the trigger according to the input, model, or target condition~\cite{salem2022dynamic,kwon2022multi}. These approaches complicate defenses that assume the existence of a single fixed trigger. Nevertheless, they are still generally designed to induce consistent attacker-specified behavior and are evaluated primarily under high-ASR conditions.

Our work examines a complementary dimension of attack stealth. Rather than further concealing the trigger pattern, we deliberately weaken the frequency with which the trigger induces the target prediction. This allows us to isolate the effect of observable backdoor strength while retaining the original trigger family and attack configuration.
\subsection{Backdoor Detection and Trigger Reconstruction}

Backdoor defenses exploit a range of signals to identify compromised models or inputs. Input-level methods analyze prediction behavior or transform suspicious inputs~\cite{gao2019strip,doan2020februus,guo2023scale}, while data- and representation-based approaches identify anomalies in training samples, learned features, or model activations~\cite{chen2018detecting,tran2018spectral,krauss2024verify,zheng2021topological,pan2023asset,qi2023proactive,liu2019abs,fu2023freeeagle}. Meta-classification approaches instead learn to distinguish clean and backdoored models from collections of reference models~\cite{xu2021detecting,kolouri2020universal}.

Another major line of work focuses on trigger reconstruction. Neural Cleanse searches for anomalously small triggers associated with candidate target classes~\cite{wang2019neural}, while subsequent methods improve trigger inversion or operate under more restricted access settings~\cite{wang2020practical,tao2022better,shen2021backdoor,guo2021aeva,dong2021black}. DeBackdoor further generalizes trigger recovery by optimizing a continuous approximation of ASR (cASR) over a broad trigger search space~\cite{popovic2025debackdoor}.

Despite their methodological differences, these defenses rely on observable behavioral or representational signals produced by the backdoor. Our work examines how deliberately reducing ASR affects these signals across four representative paradigms: STRIP, Neural Cleanse, FreeEagle, and DeBackdoor.

\subsection{Adversarial Optimization and Attacker Utility}

Gradient-based adversarial attacks manipulate an input at inference time to induce a desired prediction. Early studies demonstrated the sensitivity of neural networks to carefully constructed perturbations~\cite{szegedy2014intriguing,goodfellow2015explaining}, while later work introduced stronger optimization procedures such as projected gradient descent and the Carlini--Wagner attack~\cite{madry2018towards,carlini2017towards}.

Adversarial attacks and backdoor attacks differ in how malicious behavior is established. An adversarial attack constructs an input-specific perturbation through inference-time optimization, whereas a backdoor implants persistent attacker-induced behavior during training. In our work, gradient-based optimization is not treated as an alternative backdoor attack or as an evaluation of adversarial robustness. Instead, it provides a reference for measuring the effort required to induce the attacker-specified target prediction.

Specifically, we compare the optimization effort required for inputs evaluated on clean and low-ASR models. If triggered inputs evaluated on a low-ASR model reach the target prediction with less optimization effort than comparable inputs evaluated on a clean model, this provides evidence of residual attacker-induced target-class bias. This evaluation complements global ASR by measuring whether the weakened model remains easier for an attacker to activate using a retained local copy.

\subsection{Positioning of Our Work}

Prior research has primarily improved backdoor stealth by concealing the trigger, changing the poisoning mechanism, or making the trigger dynamic. Conversely, existing defenses attempt to recover or detect evidence produced by trigger-induced behavior. Despite their methodological differences, both lines of research commonly evaluate backdoors in a high-ASR regime.

Our work investigates a different but complementary threat dimension: deliberate reduction of the observable trigger response. Starting from a conventional high-ASR backdoor, we use reverse training to generate controlled low-ASR variants while keeping the trigger transformation, target label, architecture, and original attack configuration unchanged. This design separates the visibility of the trigger from the frequency with which it activates the attacker-specified behavior.

We then examine two related questions: whether representative defenses remain effective as ASR decreases and whether the resulting models retain measurable attacker-useful target-class bias. In contrast to prior stealthy-trigger attacks, our objective is not to hide the trigger pattern itself, but to weaken the behavioral signal on which existing defenses rely. Low-ASR backdoors therefore expose a limitation of evaluation paradigms that treat strong and consistent trigger activation as a necessary property of an operational backdoor.

%% file: sections/7_conclusion.tex
\section{Conclusion}
\label{sec:conclusion}

Backdoor defenses are largely designed and evaluated under the assumption that effective attacks exhibit high attack success rates (ASR). In this paper, we challenged this assumption by studying low-ASR backdoors, a regime in which the attacker deliberately suppresses observable trigger-induced behavior while preserving residual malicious functionality.

To systematically investigate this setting, we introduced a reverse training framework that transforms conventional high-ASR backdoors into low-ASR variants while maintaining clean model performance. Through extensive experiments across multiple datasets, architectures, trigger families, and attack types, we demonstrated that low-ASR backdoors retain measurable trigger-related behavior despite exhibiting substantially weaker observable signals.

Our results show that existing defenses, including Neural Cleanse, STRIP, FreeEagle, and DeBackdoor consistently fail in the low-ASR regime. This failure extends beyond static patch triggers to dynamic and input-dependent attacks such as WaNet and LIRA. Furthermore, optimization-based activation experiments reveal that low-ASR backdoors remain operationally useful to an adversary even after their global attack success rates have been significantly reduced.

Taken together, these findings suggest that attack success rate should not be treated as a direct proxy for backdoor risk. A backdoor can simultaneously evade detection, exhibit low observable activation rates, and retain exploitable target-class bias. We hope this work motivates future defenses that focus on identifying latent backdoor representations rather than relying solely on strong trigger-induced behavior.

%% file: appendix/0_appendix.tex
\section{Additional Spatial Trigger Analysis}
\label{app:spatial_trigger_analysis}

\begin{figure}[H]
\centering

\begin{subfigure}{0.558\linewidth}
    \centering
    \includegraphics[width=\linewidth]{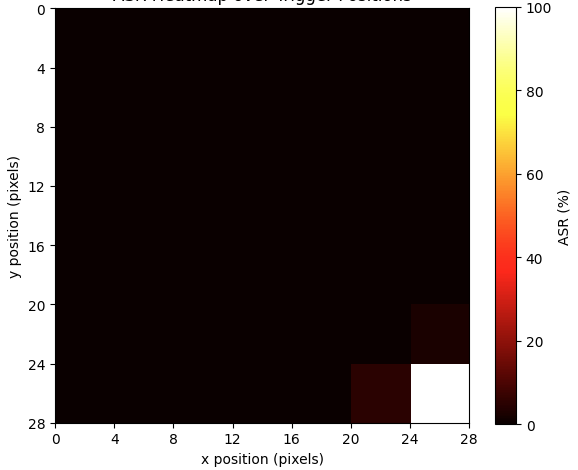}
    \caption{High-ASR}
\end{subfigure}
\hfill
\begin{subfigure}{0.58\linewidth}
    \centering
    \includegraphics[width=\linewidth]{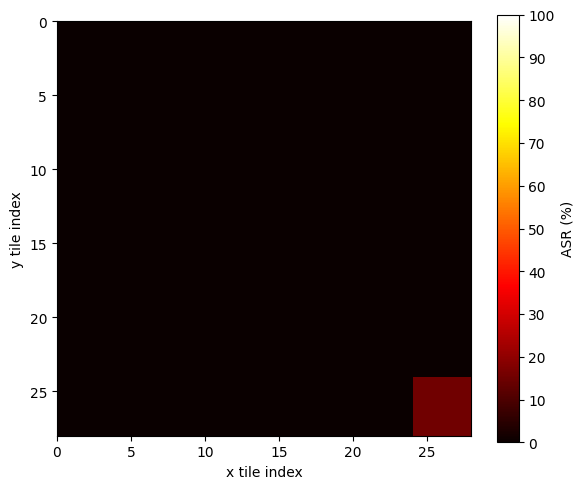}
    \caption{Low-ASR}
\end{subfigure}

\caption{ASR heatmaps obtained by inserting the trigger at different spatial locations on MNIST. In the high-ASR model, trigger effectiveness is highly localized around the original insertion region. Reverse training substantially suppresses trigger effectiveness, resulting in a much weaker spatial response in the low-ASR model.}
\label{fig:mnist_trigger_heatmaps_appendix}

\end{figure}

To further investigate the effect of reverse training on trigger robustness, we analyze the spatial behavior of triggers on MNIST. Figure~\ref{fig:mnist_trigger_heatmaps_appendix} shows ASR heatmaps obtained by inserting the trigger at different image locations for both high-ASR and low-ASR models.

Unlike CIFAR-10, which exhibits broader spatial tolerance, MNIST demonstrates highly localized trigger behavior. The trigger remains most effective near the original insertion location and rapidly loses effectiveness as it is moved away from this region. After reverse training, the overall activation strength is significantly reduced, although a small residual spatial footprint remains visible. These observations further support the conclusion that reverse training weakens the trigger--target association without completely eliminating the underlying trigger representation.

\section{Additional STRIP Analysis}
\label{app:strip_analysis}

\begin{figure*}[t]
    \centering
    \includegraphics[width=\linewidth]{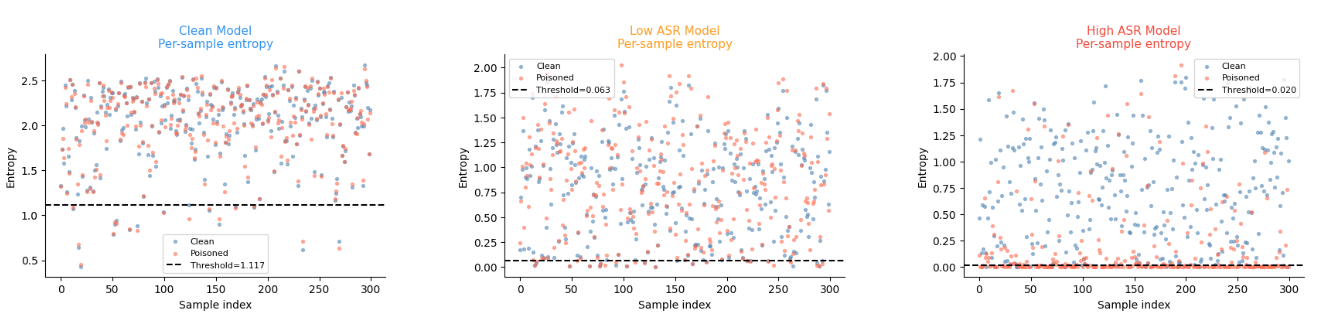}
    \caption{Per-sample entropy values under STRIP perturbations. High-ASR poisoned samples form a separable low-entropy cluster, whereas low-ASR poisoned samples overlap with clean inputs.}
    \label{fig:strip_scatter_appendix}
\end{figure*}

To provide a more detailed view of STRIP behavior, Figure~\ref{fig:strip_scatter_appendix} reports per-sample entropy values under STRIP perturbations for clean, high-ASR, and low-ASR models. In high-ASR models, poisoned samples consistently produce low entropy values and form a clearly separable cluster below the detection threshold. This behavior reflects the strong trigger--target association learned during backdoor training, which causes predictions to remain highly stable even under substantial input perturbations. As a result, STRIP can reliably distinguish poisoned inputs from clean inputs. In contrast, low-ASR models exhibit a fundamentally different pattern. Poisoned samples no longer form a distinct low-entropy cluster and instead become intermingled with clean samples across the entropy range. The resulting overlap substantially reduces the separability between clean and poisoned inputs, making reliable detection difficult. These observations provide sample-level evidence for the distributional behavior reported in Figure~\ref{fig:strip_entropy}. While the aggregate entropy distributions already indicate a collapse of separability in the low-ASR regime, Figure~\ref{fig:strip_scatter_appendix} demonstrates that this phenomenon persists at the level of individual samples. Consequently, the entropy-based signatures exploited by STRIP largely disappear once the trigger--target association is weakened through reverse training.

%% file: appendix/1_appendix.tex
\section{Extended DeBackdoor Analysis}
\label{app:debackdoor}

This appendix provides additional DeBackdoor optimization results supporting the defense evaluation in Section~\ref{sec:defense_evaluation}. We report results across MNIST, CIFAR-10, and GTSRB to show that the failure of DeBackdoor under low-ASR settings is not dataset-specific. Across datasets, high-ASR models produce a dominant target-class signal, whereas low-ASR models produce weaker, more distributed, or unstable cASR profiles.

\subsection{MNIST}
\label{app:debackdoor_mnist}

Table~\ref{tab:mnist_debackdoor} reports the highest cASR recovered by DeBackdoor under different trigger configurations on MNIST. The results show that DeBackdoor optimization is sensitive to the strength and structure of the trigger. In low-ASR settings, the recovered signal becomes less stable and no longer provides reliable evidence of a strong trigger--target mapping.

\begin{table}[t]
\centering
\small
\caption{DeBackdoor optimization results on MNIST.}
\label{tab:mnist_debackdoor}
\begin{tabular}{lcc}
\toprule
\textbf{Configuration} & \textbf{Highest cASR} & \textbf{Runtime (s)} \\
\midrule
Static White, Low-ASR & 0.2584 & 27.99 \\
Static White, 100\% ASR    & 0.9867 & 46.67 \\
Dynamic White, Low-ASR & 0.2205 & 41.73 \\
Dynamic White, 100\% ASR   & 0.9200 & 42.42 \\
Static QR, Low-ASR     & 0.2697 & 42.42 \\
Static QR, 100\% ASR       & 1.0000 & 41.47 \\
Dynamic QR, Low-ASR    & 0.2793 & 46.48 \\
Dynamic QR, 100\% ASR      & 1.0000 & 45.06 \\
\bottomrule
\end{tabular}
\end{table}

Although some MNIST low-ASR configurations yield high cASR values, these results are unstable across trigger types and do not consistently correspond to a reliable target-class signal. This suggests that DeBackdoor may overfit to simple input structures or recover spurious trigger-like patterns when the true trigger--target association is weak.

\subsection{CIFAR-10}
\label{app:debackdoor_cifar}

Tables~\ref{tab:cifar_static_casr} and~\ref{tab:cifar_dynamic_casr} report class-wise cASR values for CIFAR-10 under static and dynamic patch-trigger configurations. In the high-ASR setting, the attacker-specified target class exhibits a clear peak. In the low-ASR setting, this dominance disappears and cASR values become distributed across multiple classes.

\begin{table}[t]
\centering
\small
\setlength{\tabcolsep}{4pt}
\caption{Class-wise cASR for static patch triggers on CIFAR-10.}
\label{tab:cifar_static_casr}
\begin{tabular}{ccccc}
\toprule
\textbf{Target} &
\textbf{100\% White} &
\textbf{Low White} &
\textbf{100\% QR} &
\textbf{Low QR} \\
\midrule
0 & 0.3156 & 0.2340 & 0.2834 & 0.2157 \\
1 & \textbf{0.9196} & 0.1535 & \textbf{0.9909} & 0.1190 \\
2 & 0.4622 & 0.2067 & 0.3372 & 0.3757 \\
3 & 0.2569 & \textbf{0.4039} & 0.3472 & 0.3747 \\
4 & 0.3024 & 0.3262 & 0.2394 & 0.3005 \\
5 & 0.4032 & 0.2203 & 0.4760 & 0.2839 \\
6 & 0.1793 & 0.2975 & 0.2725 & 0.2748 \\
7 & 0.2288 & 0.2464 & 0.3322 & 0.2429 \\
8 & 0.2382 & 0.1969 & 0.1852 & 0.1240 \\
9 & 0.1969 & 0.2091 & 0.1501 & 0.1463 \\
\bottomrule
\end{tabular}
\end{table}

\begin{table}[t]
\centering
\setlength{\tabcolsep}{4pt}
\small
\caption{Class-wise cASR for dynamic patch triggers on CIFAR-10.}
\label{tab:cifar_dynamic_casr}
\begin{tabular}{ccccc}
\toprule
\textbf{Target} &
\textbf{100\% White} &
\textbf{Low White} &
\textbf{100\% QR} &
\textbf{Low QR} \\
\midrule
0 & 0.2243 & 0.2159 & 0.3055 & 0.2742 \\
1 & \textbf{0.8889} & 0.2138 & \textbf{0.9304} & 0.1133 \\
2 & 0.2116 & 0.3656 & 0.3750 & 0.3021 \\
3 & 0.4668 & \textbf{0.4868} & 0.5767 & \textbf{0.6006} \\
4 & 0.2785 & 0.3019 & 0.2752 & 0.2470 \\
5 & 0.5946 & 0.1787 & 0.2654 & 0.1719 \\
6 & 0.1601 & 0.2527 & 0.2449 & 0.2624 \\
7 & 0.2983 & 0.2661 & 0.3235 & 0.2788 \\
8 & 0.3329 & 0.2441 & 0.2389 & 0.2176 \\
9 & 0.1759 & 0.1278 & 0.1277 & 0.0842 \\
\bottomrule
\end{tabular}
\end{table}

In high-ASR configurations, the target label consistently produces the highest cASR, indicating that the optimization process recovers a strong trigger representation. In low-ASR configurations, non-target classes often achieve comparable or higher cASR values than the intended target class. This indicates that the optimization landscape becomes ambiguous once the trigger--target association is weakened.

\subsection{GTSRB}
\label{app:debackdoor_gtsrb}

Table~\ref{tab:gtsrb_vit_casr} summarizes DeBackdoor optimization results on GTSRB using a ViT model. The high-ASR configurations produce strong recovered cASR values, whereas low-ASR configurations produce substantially weaker and less distinctive responses.

\begin{table}[t]
\centering
\setlength{\tabcolsep}{4pt}
\small
\caption{DeBackdoor optimization results on GTSRB using ViT.}
\label{tab:gtsrb_vit_casr}
\begin{tabular}{lcc}
\toprule
\textbf{Configuration} & \textbf{Highest cASR} & \textbf{Runtime (s)} \\
\midrule
Static White, 100\% ASR  & 1.000 & 58.34 \\
Static White, Low-ASR    & 0.421 & 60.12 \\
Dynamic White, 100\% ASR & 1.000 & 62.45 \\
Dynamic White, Low-ASR   & 0.398 & 61.88 \\
Static QR, 100\% ASR     & 1.000 & 63.27 \\
Static QR, Low-ASR       & 0.420 & 64.11 \\
Dynamic QR, 100\% ASR    & 1.000 & 65.02 \\
Dynamic QR, Low-ASR      & 0.476 & 64.55 \\
\bottomrule
\end{tabular}
\end{table}

Table~\ref{tab:gtsrb_blended_casr} reports class-wise cASR values for blended backdoor attacks on GTSRB. The high-ASR model exhibits a dominant response at the attacker-specified target class, whereas the low-ASR model produces a more diffuse class-wise cASR profile.

\begin{table}[t]
\centering
\setlength{\tabcolsep}{2pt}
\small
\caption{Class-wise cASR values for blended attacks on GTSRB.}
\label{tab:gtsrb_blended_casr}
\begin{tabular}{c c c || c c c}
\toprule
\textbf{Label} & \textbf{High-ASR} & \textbf{Low-ASR} &
\textbf{Label} & \textbf{High-ASR} & \textbf{Low-ASR} \\
\midrule
0  & 0.03 & 0.0483 & 22 & 0.03 & 0.0356 \\
1  & 0.04 & 0.1637 & 23 & 0.03 & 0.0378 \\
2  & 0.03 & 0.0379 & 24 & 0.03 & 0.0378 \\
3  & 0.03 & 0.0621 & 25 & 0.03 & 0.0378 \\
4  & \textbf{0.99} & \textbf{0.4992} & 26 & 0.03 & 0.3199 \\
5  & 0.03 & 0.0356 & 27 & 0.03 & 0.0338 \\
6  & 0.03 & 0.0378 & 28 & 0.03 & 0.0378 \\
7  & 0.03 & 0.0379 & 29 & 0.03 & 0.0756 \\
8  & 0.03 & 0.0378 & 30 & 0.03 & 0.2626 \\
9  & 0.03 & 0.2496 & 31 & 0.03 & 0.0328 \\
10 & 0.03 & 0.2782 & 32 & 0.03 & 0.0378 \\
11 & 0.03 & 0.0378 & 33 & 0.03 & 0.0378 \\
12 & 0.03 & 0.0378 & 34 & 0.03 & 0.0378 \\
13 & 0.03 & 0.4284 & 35 & 0.03 & 0.0378 \\
14 & 0.03 & 0.0378 & 36 & 0.03 & 0.0378 \\
15 & 0.03 & 0.0259 & 37 & 0.03 & 0.0378 \\
16 & 0.03 & 0.0378 & 38 & 0.03 & 0.3252 \\
17 & 0.03 & 0.0378 & 39 & 0.03 & 0.0940 \\
18 & 0.03 & 0.0378 & 40 & 0.03 & 0.0378 \\
19 & 0.03 & 0.0378 & 41 & 0.03 & 0.0091 \\
20 & 0.03 & 0.0378 & 42 & 0.03 & 0.0622 \\
21 & 0.03 & 0.0378 &    &      &        \\
\bottomrule
\end{tabular}
\end{table}

The GTSRB results follow the same trend observed on CIFAR-10. High-ASR models yield a clear target-class signal, while low-ASR models produce weaker and more distributed responses. This confirms that DeBackdoor's effectiveness is strongly tied to the presence of a dominant trigger representation.

\subsection{Cross-Dataset Summary}
\label{app:debackdoor_summary}

Across MNIST, CIFAR-10, and GTSRB, the same qualitative pattern emerges. When the backdoor signal is strong, DeBackdoor can recover a trigger that produces a dominant target-class response. When reverse training weakens the trigger--target association, the recovered cASR values become unstable, diffuse, or comparable across multiple classes.

These results support the main finding of Section~\ref{sec:debackdoor}: DeBackdoor measures the strength of the recovered trigger rather than the existence of a backdoor itself. Consequently, low-ASR backdoors evade detection by suppressing the optimization signal required for reliable trigger recovery.

%% file: appendix/2_appendix.tex
\section{Dynamic and Complex Backdoor Attacks: Extended Analysis}
\label{app:dynamic_backdoors}

This appendix provides additional optimization results for dynamic and input-dependent backdoor attacks. We analyze LIRA and WaNet to examine how reducing ASR affects optimization-based trigger recovery. Unlike static patch triggers, these attacks use learned or spatially varying transformations, making the recovered cASR distribution less concentrated around a single fixed trigger pattern.

\subsection{LIRA on CIFAR-10}
\label{app:lira_cifar}

Table~\ref{tab:lira_cifar_appendix} reports class-wise cASR values for LIRA on CIFAR-10. In the high-ASR setting, the attacker-specified target class exhibits a dominant cASR value. In the low-ASR setting, this dominance disappears and the recovered signal becomes distributed across multiple classes.

\begin{table}[t]
\centering
\small
\caption{Class-wise cASR values for LIRA on CIFAR-10.}
\label{tab:lira_cifar_appendix}
\begin{tabular}{c c c}
\toprule
\textbf{Target} & \textbf{High-ASR} & \textbf{Low-ASR} \\
\midrule
0 & \textbf{0.9767} & 0.1168 \\
1 & 0.0320 & 0.0626 \\
2 & 0.0156 & 0.1356 \\
3 & 0.0002 & 0.1061 \\
4 & 0.0023 & \textbf{0.2482} \\
5 & 0.0000 & 0.0707 \\
6 & 0.0270 & 0.1215 \\
7 & 0.0733 & 0.1906 \\
8 & 0.0312 & 0.0835 \\
9 & 0.0123 & 0.0339 \\
\bottomrule
\end{tabular}
\end{table}

The high-ASR model produces a clear target-class peak at class 0, showing that DeBackdoor can recover a strong trigger signal when the trigger--target association is dominant. In contrast, the low-ASR model no longer produces a unique target-class peak. Several non-target classes obtain comparable or larger cASR values, indicating that the recovered optimization signal is diffuse and non-discriminative.

\subsection{LIRA on MNIST}
\label{app:lira_mnist}

Table~\ref{tab:lira_mnist_appendix} reports class-wise cASR values for LIRA on MNIST. The high-ASR setting produces a clear peak at the attacker-specified target class, whereas the low-ASR setting substantially reduces the recovered target-class response.

\begin{table}[t]
\centering
\small
\caption{Class-wise cASR values for LIRA on MNIST.}
\label{tab:lira_mnist_appendix}
\begin{tabular}{c c c}
\toprule
\textbf{Target} & \textbf{High-ASR} & \textbf{Low-ASR} \\
\midrule
0 & 0.1566 & 0.1525 \\
1 & \textbf{0.9229} & \textbf{0.1889} \\
2 & 0.0809 & 0.0621 \\
3 & 0.1251 & 0.1135 \\
4 & 0.0638 & 0.0570 \\
5 & 0.0880 & 0.0632 \\
6 & 0.1020 & 0.0956 \\
7 & 0.0628 & 0.0609 \\
8 & 0.1896 & 0.1210 \\
9 & 0.1853 & 0.1819 \\
\bottomrule
\end{tabular}
\end{table}

The MNIST results show the same qualitative trend as CIFAR-10. When ASR is high, the attacker-specified class is clearly separated from the remaining classes. After ASR reduction, this separation largely disappears and the recovered cASR values become closer across classes. This weakens the evidence available to optimization-based detection methods.

\subsection{WaNet on GTSRB}
\label{app:wanet_gtsrb}

Table~\ref{tab:wanet_gtsrb_appendix} reports class-wise cASR values for WaNet on GTSRB using ResNet18. Unlike LIRA, WaNet introduces a spatial warping transformation, producing a more distributed activation pattern across classes.

\begin{table}[t]
\centering
\setlength{\tabcolsep}{2pt}
\small
\caption{Class-wise cASR values for WaNet on GTSRB using ResNet18.}
\label{tab:wanet_gtsrb_appendix}
\begin{tabular}{c c c || c c c}
\toprule
\textbf{Label} & \textbf{High-ASR} & \textbf{Low-ASR} &
\textbf{Label} & \textbf{High-ASR} & \textbf{Low-ASR} \\
\midrule
0  & \textbf{0.9999} & 0.3791 & 22 & 0.0613 & 0.1280 \\
1  & 0.3027 & 0.1516 & 23 & 0.7544 & 0.7719 \\
2  & 0.3066 & 0.2831 & 24 & 0.3377 & 0.1853 \\
3  & 0.3893 & 0.5244 & 25 & 0.8850 & 0.9098 \\
4  & 0.3110 & 0.4837 & 26 & 0.2049 & 0.6299 \\
5  & 0.3397 & 0.2773 & 27 & 0.0899 & 0.0313 \\
6  & 0.1812 & 0.1563 & 28 & 0.8680 & 0.1100 \\
7  & 0.3611 & 0.2464 & 29 & 0.2628 & \textbf{0.9525} \\
8  & 0.4806 & 0.5889 & 30 & 0.2822 & 0.3105 \\
9  & 0.6910 & 0.7811 & 31 & 0.1249 & 0.6127 \\
10 & 0.5196 & 0.4883 & 32 & 0.5192 & 0.2434 \\
11 & 0.5555 & 0.6591 & 33 & 0.5808 & 0.9304 \\
12 & 0.6487 & 0.4493 & 34 & 0.1158 & 0.1932 \\
13 & 0.9991 & 0.9317 & 35 & 0.5512 & 0.4275 \\
14 & 0.0595 & 0.1448 & 36 & 0.0623 & 0.1208 \\
15 & 0.5134 & 0.2618 & 37 & 0.0466 & 0.1822 \\
16 & 0.1218 & 0.2439 & 38 & 0.8684 & 0.3819 \\
17 & 0.4687 & 0.8084 & 39 & 0.3814 & 0.6688 \\
18 & 0.5696 & 0.5196 & 40 & 0.6015 & 0.2446 \\
19 & 0.0779 & 0.0070 & 41 & 0.1534 & 0.1503 \\
20 & 0.4197 & 0.9449 & 42 & 0.0617 & 0.1250 \\
21 & 0.0595 & 0.1027 &    &        &        \\
\bottomrule
\end{tabular}
\end{table}

The WaNet results reveal a more complex optimization landscape than static or LIRA-style triggers. In the high-ASR setting, several classes exhibit elevated cASR values, indicating that spatial warping can produce broader class-wise activation patterns. In the low-ASR setting, the response becomes even more distributed: multiple non-target classes achieve high cASR values, and no single class provides a clean, uniquely identifiable trigger signature.

This behavior challenges the single-peak assumption underlying cASR-based detection. When the backdoor signal is spatially distributed or weakened through reverse training, DeBackdoor may recover ambiguous class-wise responses rather than a dominant target-class trigger. This supports the main paper's finding that optimization-based defenses measure the strength and separability of the recovered trigger signal, not the mere existence of a backdoor.

\subsection{Summary}
\label{app:dynamic_backdoors_summary}

Across LIRA and WaNet, reducing ASR consistently weakens the dominance of the recovered target-class signal. For LIRA, the high-ASR model yields a clear target-class peak, while the low-ASR model produces a flatter cASR profile. For WaNet, the spatially distributed trigger produces multi-class activation patterns, making the optimization output even less separable. These results support the conclusion that low-ASR dynamic backdoors evade detection by suppressing the strong optimization signal required for reliable trigger recovery.